\documentclass[%
 reprint,
 amsmath,amssymb,
aps,
]{revtex4-2}

\usepackage{graphicx}
\usepackage{dcolumn}
\usepackage{bm}

\usepackage{mystyle}

\begin{document}

\preprint{APS/123-QED}

\title{Magnetism of the Bilayer Wigner Crystal}
\author{Ilya Esterlis}
\author{Dmitry Zverevich}
\author{Zekun Zhuang}
\author{Alex Levchenko}
\affiliation{Department of Physics, University of Wisconsin-Madison, Madison, Wisconsin 53706, USA}

\date{\today}

\begin{abstract}
The multiple-spin exchange frequencies of the bilayer Wigner crystal are determined by the semiclassical method, which is asymptotically exact in the limit of dilute electron densities. The evolution of the exchange frequencies with interlayer distance -- as the crystal undergoes a sequence of structural transitions -- leads to both ferromagnetic and multi-sublattice antiferromagnetic phases. Extrapolation of the results to higher density suggests a rich magnetic phase diagram of the bilayer Wigner crystal, including the possibility of spin-nematic and valence-bond solid phases. The bilayer crystal is stable to higher electron densities than the monolayer, leading to enhanced magnetic energy scales. Our estimates of the exchange energies suggest some of the magnetic phases may be accessible in recently discovered bilayer Wigner crystals in a transition-metal dichalcogenide system.
\end{abstract}

\maketitle

\section{Introduction}

Bilayer electronic systems, consisting of two Coulomb-coupled two-dimensional (2D) layers of electrons separated by a distance $d$, exhibit a variety of novel phenomena not possible in a single monolayer. The example of a bilayer two-dimensional electron gas (2DEG) -- realized in semiconductor-based systems such as coupled quantum wells and bilayers of certain atomically-thin 2D materials --  is of particular interest owing to its relative simplicity and rich phase diagram. Here we will be interested in the behavior of such a system when the electron density is low and the electron-electron interactions are correspondingly strong. While a monolayer 2DEG freezes into a triangular lattice Wigner crystal (WC) \cite{wigner1934,bonsall1977} at low electron density, the Coulomb interaction between two dilute layers in a bilayer system leads to a variety of new bilayer WC (BLWC) phases. The resulting crystal structure and magnetic order depend on the ratio of $d$ to the average interparticle spacing $a$ within a layer \cite{Vilk1984,*Vilk1985,Falko1994,Narsimhan1995,Esfarjani1995,goldoni1996,goldoni1997}.

In the present paper we explore the multiple-spin exchange processes and magnetism that arises in such BLWCs. In the dilute limit, where controlled calculations of the exchange couplings are possible, we find a variety of magnetic phases that can be realized as a function of $d/a$. The extrapolation of our results to higher density suggests a rich magnetic phase diagram of the bilayer 2DEG, including exotic quantum phases. 

The Hamiltonian of the bilayer 2DEG is   
    \be
    \begin{aligned}
    H = \sum_{\ell, i}\frac{\bfp_{\ell,i}^2}{2m} &+ \frac 12\sum_{i\neq j} \frac{e^2}{\epsilon|\bfr_i^{(1)} - \bfr_j^{(1)}|} + \frac 12\sum_{i\neq j} \frac{e^2}{\epsilon|\bfr_i^{(2)} - \bfr_j^{(2)}|} \\
    &\quad  + \sum_{ij} \frac{e^2}{\epsilon \sqrt{|\bfr_i^{(1)} - \bfr_j^{(2)}|^2 + d^2}}.
    \end{aligned}
    \label{eq:biH}
    \ee
Here $\ell = 1,2$ is the layer index and indices $i,j$ label electrons within a layer, i.e., $\bfr_i^{(\ell)}$ and $\bfp_{\ell,i}$ refer to the respective position and momentum of electron $i$ in  layer $\ell$; $m$ is the effective electron (or hole) mass; and $\epsilon$ is the dielectric constant of the environment. Implicit is the presence of a neutralizing background for each layer. We will consider the situation in which the electron densities in the layers are equal, $n_1 = n_2 \equiv n$. In this case the ground-state properties of the system depend on two dimensionless parameters: the ratio $d/a$ and the usual electron gas parameter $r_s = a/a_B$, where $a$ is related to the density according to $n \pi a^2 = 1$ and $a_B = \hbar^2\epsilon/ me^2$ is the effective Bohr radius of the host semiconductor. The dilute limit, corresponding to large $r_s$, is the limit of strong intralayer interactions. The strength of interlayer interactions is determined by $d/a$. 

\begin{figure*}[t!]
\includegraphics[width=\textwidth]{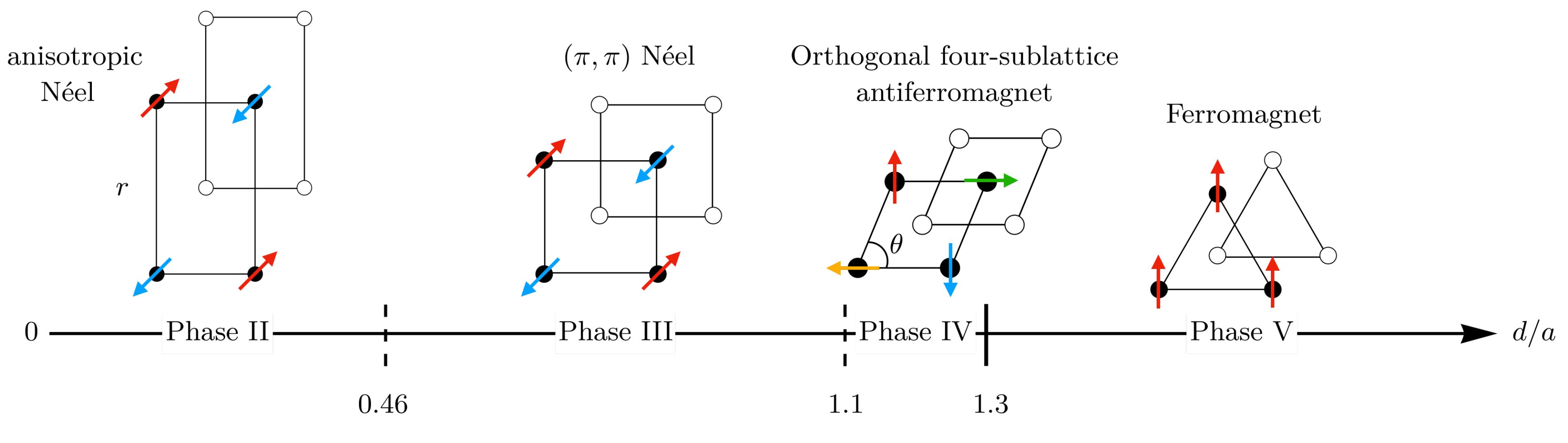}
\caption{Magnetic phase diagram of the bilayer WC in the $r_s \to \infty$ limit as a function of interlayer distance $d/a$. For clarity, the magnetic order is shown in only one of the two layers. In the staggered triangular crystal (Phase V) the bilayer WC is a ferromagnet. There is a first-order transition when $d/a \approx 1.3$ to Phase IV -- a ``soft" rhombic phase in which the angle $\theta$ between the primitive vectors varies continuously with $d/a$ -- possessing orthogonal four-sublattice antiferromagnetic order. A continuous transition occurs at $d/a \approx 1.1$ to the staggered square crystal (Phase III), which has a narrow range of four-sublattice antiferromagnetic order (not shown), transitioning to a $(\pi,\pi)$ Néel state when $d/a\approx1.0$. When $d/a\approx 0.46$ there is a transition to the soft rectangular crystal (Phase II), in which the aspect ratio $r$ varies continuously with $d/a$, and the system is an anisotropic Néel antiferromagnet, becoming increasingly one-dimensional as $d/a \to 0$.}
\label{fig:phase_diag_1d}
\end{figure*}

The properties in the extreme dilute limit, obtained by taking $r_s \to \infty$ with $d/a$ fixed, are well understood. In this regime the system is purely classical and the ground state crystal structure is determined by minimizing the electrostatic Coulomb energy \cite{Vilk1984,Vilk1985,Falko1994,Narsimhan1995,Esfarjani1995,goldoni1996}. We briefly summarize the results: at infinite interlayer separation the layers form decoupled triangular lattices. As $d/a$ is decreased, the two triangular lattices lock into a staggered triangular lattice structure. Throughout the paper we will follow the nomenclature of Ref.~\cite{goldoni1996} in labeling the various structural phases, where the staggered triangular crystal was called Phase V. Upon further decreasing $d/a$ there is a first-order transition to a staggered rhombic lattice (Phase IV). The angle $\theta$ between the primitive vectors continuously changes with decreasing $d/a$, until the lattice locks into a staggered square structure (Phase III).  The staggered square structure remains stable down to a critical value of $d/a$, beyond which the aspect ratio $r$ changes continuously and the system forms a staggered rectangular lattice (Phase II). This evolution continues to $d/a=0$, at which point the system becomes a single triangular lattice with twice the original density \footnote{In Ref.~\cite{goldoni1996} it was concluded that there is a non-zero range near $d/a=0$ where a single-component triangular lattice is stable (Phase I). It was later shown, however, that the single-component triangular lattice is stable only at $d/a=0$ \cite{samaj2012}, with the aspect ratio in Phase II evolving continuously down to $d/a = 0$.}. The evolution of BLWC structures is summarized in Fig.~\ref{fig:phase_diag_1d}. 

Away from the classical $r_s \to \infty$ limit, there have been attempts to determine aspects of the zero-temperature phase diagram as a function of $d/a$ and $r_s$, the primary focusing being the quantum melting transition. This includes a perturbative RPA-like treatment of the interlayer interaction \cite{swierkowski1991}, a DFT study \cite{goldoni1997}, and quantum Monte Carlo simulation \cite{rapisarda1996,*rapisarda1998}. The most important conclusion common to all these approaches is the enhanced stability of the bilayer crystal phase relative to the monolayer. While the quantum melting of the monolayer WC occurs at $r_s \approx 35$ \cite{Tanatar1989,Attaccalite2002,Drummond2009,azadi2024}, the bilayer crystal was predicted to be stable down to $r_s \approx 20$ at intermediate values of $d/a$ \cite{rapisarda1998}. 
Such an enhanced stability of the BLWC is expected, owing to the Coulomb energy gained in the crystal phase from commensurate locking of the lattices in the two layers. As we will elaborate upon below, the stability of the BLWC to higher density (smaller $r_s$) has a number of interesting consequences. In particular, the energy scales associated with magnetism of the electron crystal -- which is the main focus of this paper -- increase with increasing density, and the enhanced stability therefore implies higher temperature scales at which interesting magnetic phenomena could be observed.

Perhaps most intriguingly, the existence and enhanced stability of the BLWC were recently confirmed experimentally in a bilayer transition-metal dichalcogenide system \cite{Zhou2021}. Optical probes indicated formation of insulating states at precisely those density ratios between layers where one would expect commensurate stacking of triangular crystals. The insulating features persisted to significantly higher densities ($r_s \approx 10$) and temperatures ($T\approx 40 ~\mathrm K$) than those observed in the corresponding monolayer system \cite{Smolenski2021}. These developments call for a theoretical analysis of the magnetism that could be realized in such a BLWC.

\section{Approach}

As described above, in the dilute limit $r_s \to \infty$, the Coulomb energy dominates in the Hamiltonian \eqref{eq:biH}, leading to the formation of various classical BLWCs. In the classical approximation, where the kinetic energy is neglected, the electrons forming the BLWC are effectively distinguishable particles and all possible spin states are degenerate. The degeneracy is lifted by the kinetic energy, which allows for quantum tunneling between classical configurations related by permutation of electrons. This leads to an effective ``exchange" Hamiltonian in terms of permutation operators acting on the spin degrees of freedom
\cite{thouless1965,roger1983,roger1984,katano2000,chakravarty1999,voelker2001}:
    \be
    H_\text{ex} = \sum_P (-1)^{P} J_P \hat P.
    \label{eq:Hex}
    \ee
The sum is taken over cyclic permutations $P$ of electrons,  with $\hat P$ the associated permutation operator, and the exchange coupling $J_P$ is one-half the energy splitting associated to tunneling between two electronic configurations related by permutation $P$. We adopt the convention that $J_P > 0$, the sign of each term in the sum being fixed by the parity of the permutation $(-1)^{P}$, which, as shown originally by Thouless \cite{thouless1965,roger1983}, is positive for exchanges involving an even number of particles and negative for exchanges involving an odd number of particles. The exchange Hamiltonian \eqref{eq:Hex} can be brought into a generalized Heisenberg form by noting that each cyclic permutation $P$ may be decomposed as a product of transpositions, which are in turn expressed in terms of spin operators $\hat P_{ij} = 2(\bfS_i \cdot \bfS_j) + 1/2$. Concrete examples of such spin Hamiltonians will be given later.

We will work in the limit of vanishing interlayer tunneling and therefore only consider processes in which particles are exchanged in a single layer, while particles in the opposite layer adjust their positions without exchanging. We refer to them as the ``active" and ``passive" layers, respectively. With equal layer densities and vanishing interlayer tunneling the exchange Hamiltonian \eqref{eq:Hex} is the sum of two equivalent decoupled magnetic Hamiltonians, one for each layer. 

Utilizing the semiclassical instanton method  \cite{Coleman1979}, the exchange couplings $J_P$ are found to be \cite{hirashima2001,voelker2001}
    \be
    J_P = \frac{e^{2}}{\epsilon a_B} \frac{A_P}{r_s^{5/4}}\sqrt{\frac{S_P}{2\pi}} \exp\left(-\sqrt r_s S_P\right).
    \label{eq:J}
    \ee
Here $S_P$ is the Euclidean action evaluated along the classical imaginary-time trajectory that affects the permutation $P$. The prefactor $A_P$ is related to the ``fluctuation determinant", obtained by integrating over the Gaussian fluctuations about the classical path. The factor $e^2/\epsilon a_B$ is the effective Hartree energy unit for the 2DEG and sets the overall energy scale. When there is more than one classical path for a given multiparticle permutation the exchanges associated with different paths should be added. The reliability of the semiclassical approximation for the exchange couplings in the dilute regime has been established by comparison with path-integral Monte Carlo simulations of the monolayer WC \cite{Bernu2001}.

In the $r_s \to \infty$ limit, the dominant exchange is that with the smallest action. In the case of the monolayer WC, the three-particle exchange $J_3$ has been shown to have the smallest action \cite{roger1984,katano2000,chakravarty1999,voelker2001}. Hence, as $r_s \to \infty$, the monolayer WC is a ferromagnet. The situation is richer when we consider the BLWC, as we elaborate upon below. 

\section{Results}

In this section we present results for the classical imaginary-time paths associated with multi-particle exchanges in the BLWC, together with the corresponding actions $S_P$ and prefactors $A_P$. With knowledge of the actions we can draw conclusions about the magnetism of the BLWC that are asymptotically exact as $r_s \to \infty$. Our results for the magnetic phases in this limit as a function of $d/a$ \footnote{Throughout the text we quote approximate, as opposed to exact, values of $d/a$, writing $d/a \approx \ldots$. The reason is that, in the numerical calculations, we actually fix the ratio $d/a_\ell$, where $a_\ell$ is the lattice constant for a given BLWC geometry. The interparticle distance $a$ is related to $a_\ell$ according to $a=\sqrt{A_c/\pi}$, where $A_c$ is the unit cell area. In the different geometries we have $A_c = \sqrt 3 a_\ell^2/2$ (Phase V), $A_c = \sin\theta a_\ell^2$ (Phase IV), $A_c = a_\ell^2$ (Phase III), and $A_c = r a_\ell^2$ (Phase II).} are summarized in Fig.~\ref{fig:phase_diag_1d}. By also calculating the prefactors we obtain estimates of the exchange couplings \eqref{eq:J} and extend the magnetic phase diagram to finite $r_s$. 

\subsection{Classical actions and magnetism as $r_s \to \infty$} 

In the BLWC, the classical action for a given exchange process $P$ is a function of the interlayer distance, $S_P = S_P(d/a)$.
We have calculated the classical actions for various exchanges in the BLWC as a function of $d/a$ across the different bilayer geometries described above. This is done by minimizing the Euclidean action 
    \be
    S_P = \int_{\bfR_0}^{P \bfR_0} \dd R ~ \sqrt{2m[V(\bfR)- E_0]},
    \label{eq:SP}
    \ee
where $\bfR = (\bfr_1^{(1)}, \ldots, \bfr_N^{(1)}, \bfr_1^{(2)}, \ldots, \bfr_N^{(2)})$ is a $4N$-dimensional coordinate in the configuration space of the $2N$ electrons ($N$ electrons in each layer), $V(\bfR)$ is the bilayer potential energy given in \eqref{eq:biH}, and $\bfR_0$ is the ground state electronic configuration of the BLWC with $E_0$ the corresponding ground state energy, which has been subtracted for convenience. The integration is done with respect to the arc-length $\dd R$ in configuration space. In our numerical calculations we discretize the integral \eqref{eq:SP} into $M$ steps via the trapezoid rule and allow $N_\text{move}$ electrons in each layer to adjust their positions throughout the exchange, with all other electrons being fixed at their equilibrium BLWC positions. The results presented below are for $M=16$ and $N_\text{move} \approx 100$, the latter depending on the specific lattice geometry and exchange process under consideration. We believe the actions obtained with these parameters to be accurate to within 1\%; the convergence with $M$ and $N_\text{move}$ is discussed in more detail in Appendix \ref{app:conv}.
The electrostatic energy $V(\bfR)$ is efficiently computed via standard Ewald methods \cite{bonsall1977, goldoni1996}. Some additional data concerning the role of the passive layer in the exchange process can be found in Appendix \ref{app:pass}.

\subsubsection{Phase V}
\label{sec:sV}

\begin{figure}
\includegraphics[width=\columnwidth]{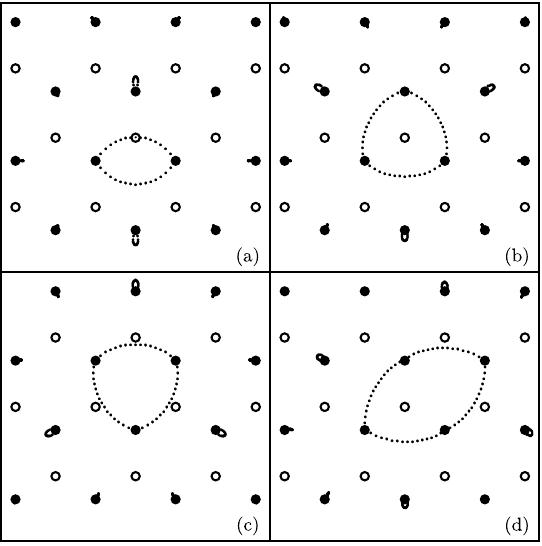}
\caption{Paths for (a) 2, (b,c) 3, and (d)  4-particle exchange in the staggered triangular geometry (Phase V) for interlayer separation $d/a \approx 1.5$. Black and white markers correspond to electrons in the active and passive layers, respectively. Dotted lines indicate trajectories of moving particles. We refer to the inequivalent 3-particle exchange paths in Panels (b) and (c) as 3 and 3$^\star$, respectively.}
\label{fig:path_V}
\end{figure}

We start from the staggered triangular phase (Phase V), which is the stable ground state for $1.3 \lesssim d/a < \infty$ \cite{goldoni1996}. As a representative example, the 2, 3, and 4-particle exchange paths obtained by minimizing Eq. \eqref{eq:SP} for $d/a \approx 1.5$ are shown in Fig.~\ref{fig:path_V}. For finite $d/a$ there are two inequivalent 3-particle exchange processes, with associated actions $S_3$ (Fig.~\ref{fig:path_V}b) and $S_3^\star$ (Fig.~\ref{fig:path_V}c), differing in the positions of the neighbors from the opposite layer. Additional data for 5 and 6-particle exchanges are reported in Appendix \ref{app:large_cyc}.

As $d/a\to \infty$, we recover the well-known result that the action for 3-particle exchange is the smallest \cite{roger1984, voelker2001, katano2000}. Upon decreasing $d/a$, we find $S_3^\star$ decreases in magnitude, while all other actions increase; see Fig.~\ref{fig:actions}.
(Interestingly, $S_3$ grows more rapidly than $S_2$, the action associated with 2-particle exchange, and there is a crossing point for $d/a \approx 1.5$). The fact that $S_3^\star$ has the smallest action throughout Phase V implies that, in the $r_s \to \infty$ limit, the 3-particle exchange $J_3^\star$ is dominant. In the dilute limit, the BLWC in Phase V is therefore a ferromagnet (Fig.~\ref{fig:phase_diag_1d}).

\subsubsection{Phase IV}
\label{sec:sIV}

\begin{figure}
\includegraphics[width=\columnwidth]{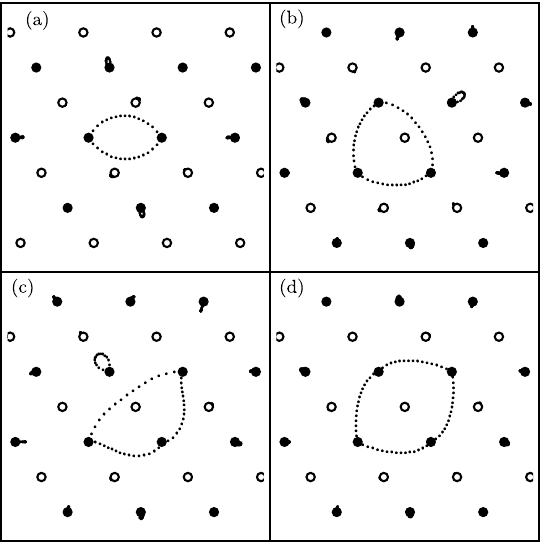}
\caption{Paths for (a) 2, (b,c) 3, and (c) 4-particle exchange in the staggered rhombic geometry (Phase IV) for interlayer separation $d/a\approx 1.2$. The 3-particle exchange path in Panel (c), which we refer to as $\widetilde 3$, involves a further neighbor exchange than that in Panel (b), the two paths becoming equivalent in Phase III.}
\label{fig:path_IV}
\end{figure}

When $d/a \approx 1.3$, there is a first-order transition from Phase V to the staggered rhombic phase (Phase IV) \cite{goldoni1996}. In this phase the angle $\theta$ between the primitive vectors of the lattice varies continuously with $d/a$. Representative 2, 3, and 4-particle exchanges in this phase are shown in Fig.~\ref{fig:path_IV} for $d/a \approx 1.2$, corresponding to $\theta \approx 73^\circ$.  We have also considered a further neighbor 3-particle exchange path, denoted $\widetilde 3$, involving the long diagonal of the rhombus (Fig.~\ref{fig:path_IV}c). We include this further neighbor path because it becomes equivalent to the more compact 3-particle exchange path involving the short diagonal (Fig.~\ref{fig:path_IV}b) in Phase III. 

Upon entering Phase IV, $S_4$, the action associate with 4-particle exchange, abruptly becomes the smallest and remains so throughout this entire structural phase; see Fig.~\ref{fig:actions}. The dominance of the 4-particle exchange in the $r_s \to \infty$ limit implies that the magnetic ground state of the BLWC in Phase IV in the dilute limit is the orthogonal four-sublattice antiferromagnet \cite{lauchli2005} (Fig.~\ref{fig:phase_diag_1d}). 

\subsubsection{Phase III}
\label{sec:sIII}

\begin{figure}
\includegraphics[width=\columnwidth]{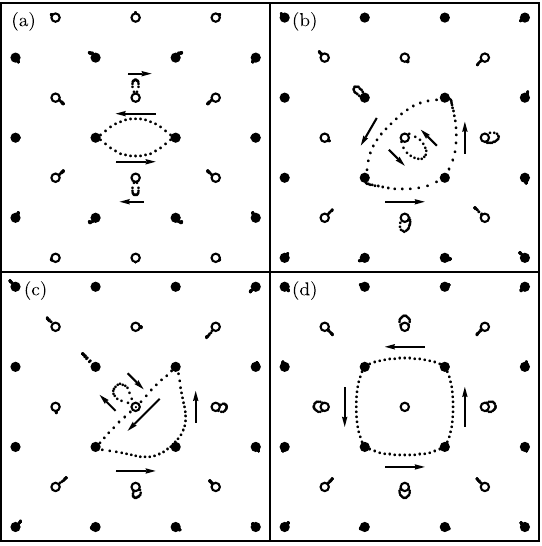}
\caption{Paths for (a) 2, (b,c) 3, and (d) 4-particle exchange in the staggered square geometry (Phase III) for interlayer separation $d/a\approx 0.53$. The exchange path in Panel (c) only appears for $d/a\lesssim 0.6$. We refer to this path as 3'.}
\label{fig:path_III}
\end{figure}

When $d/a \approx 1.1$, the angle $\theta$ between primitive vectors locks at $90^\circ$ and the structure is a staggered square lattice (Phase III) \cite{goldoni1996}. Representative 2, 3, and 4-particle exchanges in this phase are shown in Fig.~\ref{fig:path_III} for $d/a \approx 0.53$. Over much of this phase we find that there is only a single 3-particle exchange path. However, for $d/a \lesssim 0.4$,  another path, which we refer to as 3', appears (Fig.~\ref{fig:path_III}c). 

Within Phase III there is a narrow range $1.0 \lesssim d/a \lesssim 1.1$ over which $S_4$ remains the smallest action and the magnetic ground state is the 4-sublattice antiferromagnet, as in Phase IV. However, for $d/a \lesssim 1.0$,  $S_2$ becomes smaller than $S_4$. The associated 2-particle exchange $J_2$ therefore dominates as $r_s \to \infty$ and the magnetic ground state of Phase III in the dilute limit for $d/a \lesssim 1.0$ is the square lattice Néel antiferromagnet (Fig.~ \ref{fig:phase_diag_1d}).  

\subsubsection{Phase II}
\label{sec:sII}

\begin{figure}
\includegraphics[width=\columnwidth]{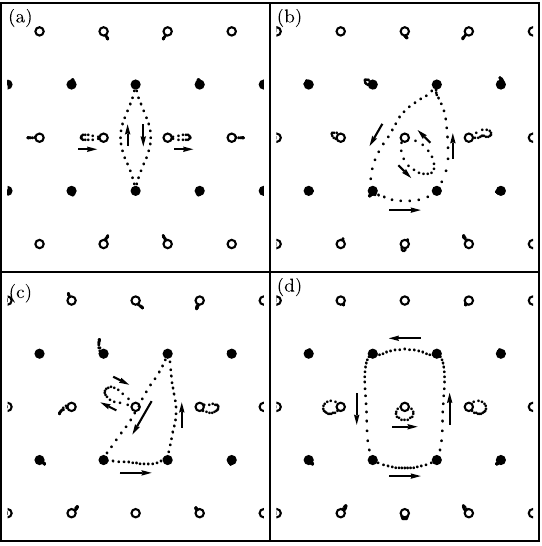}
\caption{Exchange paths in staggered rectangular geometry (Phase II) for interlayer separation $d/a \approx 0.27$. 
Arrows indicate orientation of the paths. (a) 2-particle exchange along the ``long" direction of the rectangular lattice. We denote this path $2_\perp$. The 2-particle exchange path along the ``short" direction is the same as that for the monolayer crystal (see Fig.~\ref{fig:path_V}a). (b) 3-particle exchange. (c) Additional 3-particle exchange path, referred to as $3'$.  (d) 4-particle exchange.  Similar paths in the limit $d/a = 0$ have been displayed before in Ref.~\cite{hirashima2001}.}
\label{fig:path_II}
\end{figure}

When $d/a \approx 0.46$ there is a phase transition and the aspect ratio $r$ of the square lattice begins to change continuously \cite{goldoni1996}, the lattice becoming increasingly anisotropic with decreasing $d/a$ (Phase II). As $d/a \to 0$, $r$ goes continuously from $r=1$ to $r=\sqrt 3 \approx 1.73$, thereby interpolating between the staggered square lattice and a ``one-component" triangular lattice at $d/a = 0$. Representative 2, 3, and 4-particle exchanges are shown in Fig.~\ref{fig:path_II} for $d/a \approx 0.27$, corresponding to an aspect ratio $r \approx 1.57$ \footnote{There are two symmetry-related 4-particle exchange paths in Phase II (Fig.~\ref{fig:path_II}d) and the exchange coupling $J_4$ is therefore multiplied by a factor of two.}. We note that similar paths were displayed in Ref.~\cite{hirashima2001} in the limit $d/a = 0$ \footnote{When $d/a=0$ the paths become identical to the same paths for a single-layer WC, although the meaning of 3 and 4 particle exchange is different from the single-layer case.}. There are now two 2-particle exchanges, corresponding to the horizontal and vertical directions. The latter exchange, which we refer to as $2_\perp$, has a significantly larger action as $d/a \to 0$ (Fig.~\ref{fig:actions}).  As in Phase III, there are two possible 3-particle exchange paths (Fig.~\ref{fig:path_II}b,c). 

Upon decreasing $d/a$ in Phase II, $S_2$ decreases monotonically while the actions for all other exchanges increase monotonically; see Fig.~\ref{fig:actions}. The magnetism as $r_s \to \infty$ is that of an anisotropic Heisenberg antiferromagnet with a corresponding Néel ground state.

\subsubsection{Summary: Magnetism of the dilute BLWC}

\begin{figure}
\includegraphics[width=\columnwidth]{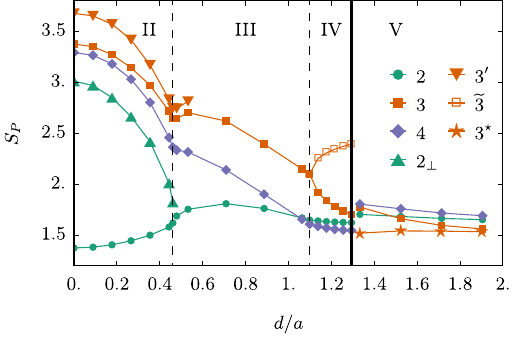}
\caption{Dimensionless actions $S_P$ for various 2-, 3-, and 4-particle exchange paths as a function of $d/a$ across the various bilayer WC geometries. Structural phase transitions are indicated by vertical lines; the transition from Phase V $\to$ Phase IV is discontinuous (solid line), while the transitions Phase IV $\to$ Phase III and Phase III $\to$ Phase II are continuous (dashed lines). Labeling of paths is explained in the main text and Figs. \ref{fig:path_V}- \ref{fig:path_II}.}
\label{fig:actions}
\end{figure}

Our results for the magnetic phases of the BLWC in the dilute limit $r_s \to \infty$ are summarized in Fig.~\ref{fig:phase_diag_1d}. Accompanying the series of structural transitions with varying $d/a$ we find a ferromagnetic phase, as well as two- and four-sublattice antiferromagnetic phases. 

For $d/a \lesssim 0.7$, $S_2$ decreases with decreasing $d/a$ while all other actions are increasing (Fig.~\ref{fig:actions}), $S_2$ becoming significantly smaller than the actions for other processes as $d/a \to 0$. In this sense, the BLWC is less magnetically frustrated than the monolayer, where various exchange processes have comparable actions.

\subsection{Prefactors}

\begin{figure}
\includegraphics[width=\columnwidth]{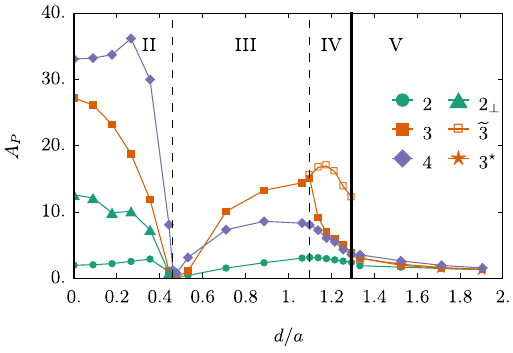}
\caption{Dimensionless amplitude prefactors $A_P$ for 2-, 3-, and 4-particle exchanges as a function of $d/a$ across the various bilayer WC geometries. The meaning of paths $2_\perp$, $\widetilde{3}$, and $3^\star$ is explained in the main text and Figures \ref{fig:path_III} and \ref{fig:path_II}. The amplitude for path $3'$ in Phases II and III is too large to be put on the present scale; e.g., $A_{3'}(d/a=0) \approx 60$.}
\label{fig:amps}
\end{figure}

At finite $r_s$, the prefactor $A_P$ in Eq.~\eqref{eq:J} plays a role in determining the dominant exchange cycle. To extend our results away from the $r_s \to \infty$ limit we have therefore also calculated the prefactors as a function of interlayer distance, $A_P = A_P(d/a)$, across the various crystal structures. In addition to yielding quantitative estimates of the exchanges at finite $r_s$, the prefactors are of interest in order to understand how the evolution of the phonon spectrum with $d/a$ influences the exchange couplings. Methods to calculate such determinants have been elaborated upon in earlier works \cite{katano2000,voelker2001,kim2024}. We have followed the approach in Refs. \cite{voelker2001,kim2024}, to which we direct the reader for further details. We note that the prefactors converge more slowly with $N_\text{move}$ than the actions, and it is therefore challenging to rigorously quantify our errors; see Appendix \ref{app:conv} for more details. 

In Fig.~\ref{fig:amps} we present our results for the prefactors $A_P$. When $d/a$ is large the prefactors for the different processes considered are all of order one, in agreement with earlier findings on the monolayer WC \cite{katano2000,voelker2001}. As $d/a$ decreases, there is a growing spread in the magnitude of the different prefactors. Some of the $A_P$ become significantly larger than unity at intermediate values of $d/a$ and as $d/a \to 0$, though these large prefactors are offset by large values of the corresponding actions (see Fig.~\ref{fig:actions}).

A striking feature of the evolution is the rapid decrease of the prefactors at the continuous transition between Phases III and II. It is possible the prefactors exactly vanish at this point, but we cannot say for certain based on the present numerical calculations. We believe this rapid suppression is due to phonon softening at the continuous III$\to$ II transition and the associated appearance of ``flat" directions in the potential landscape. This leads to large fluctuations away from the equilibrium configurations and hence a suppression of the tunneling frequency. However, this cannot be the full story, as a similar suppression of the prefactors is \textit{not} observed at the continuous transition between Phases IV and III. Though we have not pursued the question further here, we believe the behavior of the prefactors near a continuous transition warrants further investigation. We note this evolution has interesting implications, such as the possibility of re-entrant magnetic phases at finite temperature.

\subsection{Exchange couplings and magnetic Hamiltonians}

We now combine our results for the classical actions and prefactors according to Eq.~\eqref{eq:J} to obtain estimates of the various exchanges as a function of $r_s$ and $d/a$. We  also discuss the associated magnetic Hamiltonians in the different phases and speculate on the possible magnetic ground states at finite $r_s$.

\subsubsection{Phase V}

\begin{figure}
\includegraphics[]{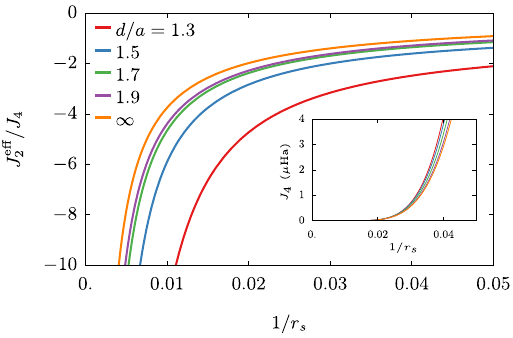}
\caption{Ratio of the effective  2-particle exchange $J_2^\text{eff}= J_2 - (J_3 + J_3^\star)$ to $J_4$ as a function of $1/r_s$ for different interlayer spacings $d/a$ in Phase V. Inset shows the magnitude of $J_4$ in micro-Hartree ($\mu$Ha) as a function of $1/r_s$.
}
\label{fig:jV}
\end{figure}

Retaining only 2, 3, and 4-particle exchange, the exchange Hamiltonian \eqref{eq:Hex} in Phase V may be expressed
    \be
    H_\text{ex}^{\text{(V)}} = J_2^\text{eff} \sum_{\includegraphics[width=0.025\textwidth]{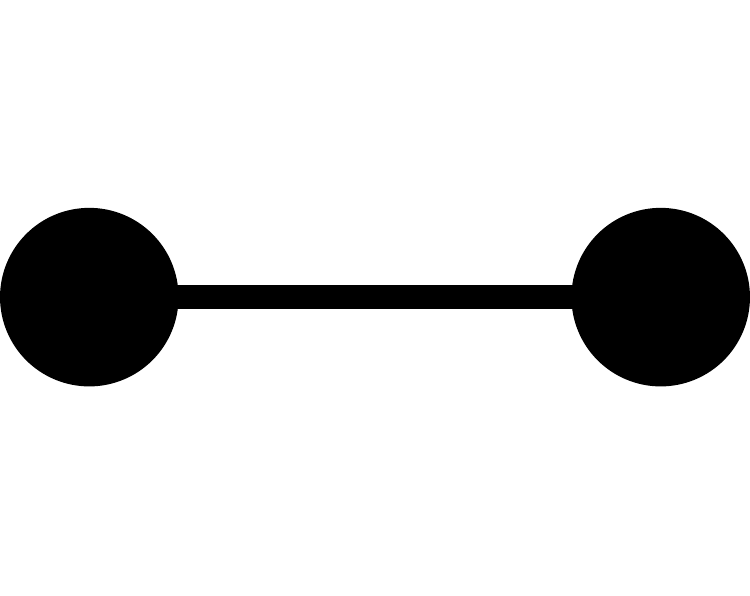}} P_2 + J_4 \sum_{\includegraphics[width=0.025\textwidth]{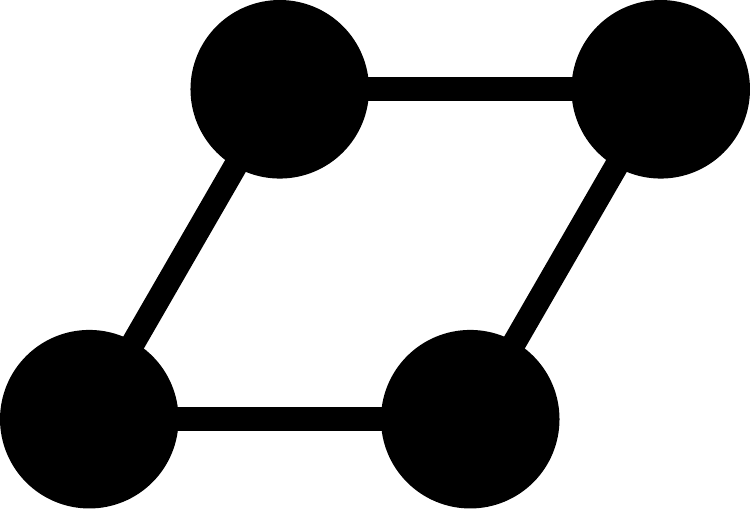}} (P_4 + P_4^{-1}). 
    \label{eq:HexV}
    \ee
Here the effective 2-particle exchange is $J_2^\text{eff} = J_2 - (J_3 + J_3^\star)$, which is obtained by reducing 3-particle exchange to a 2-particle term using the spin-1/2 identity $P_{123} + P_{321} = P_{12} + P_{23} + P_{31} -1$. The ratio $J_2^\text{eff}/J_4$ and the magnitude of $J_4$ are presented in Fig.~\ref{fig:jV}. We find $|J_2^\text{eff}|/J_4$ increases with decreasing $d/a$. 

As described in Section \ref{sec:sV}, when $r_s \to \infty$ the dominance of $J_3^\star$ leads to ferromagnetic order. Exact-diagonalization (ED) studies \cite{misguich1998,*misguich1999} on the ring-exchange Hamiltonian \eqref{eq:HexV} (which also included the 5-particle exchange term) have demonstrated that ferromagnetic order is destroyed for relatively small values of  $J_4/|J_2^\text{eff}|$ and that, for sufficiently large $J_4$, the ground state is possibly a spin-liquid, being characterized by a spin gap, short-range spin correlations, and absence of long range-order. Later exact diagonalization work that included 5 and 6-particle exchange \cite{momoi2012} found that a spin-nematic --  a magnetically ordered state that preserves time-reversal symmetry but break spin rotation symmetry -- could also be stabilized for relatively small values of $J_5$ and $J_6$.  

Rather than attempting to identify the precise magnetic ground state at finite $r_s$, which may depend sensitively on the small couplings associated with larger ring exchange processes, here we would like to highlight that the bilayer coupling simplifies the situation relative to the monolayer WC. In the monolayer case, ring exchanges involving 4 or more particles become of comparable importance to the (effective) 2-particle exchange at relatively large values of $r_s$, leading to highly frustrated magnetic interactions and the complex magnetic phase diagram described above \cite{misguich1998,misguich1999,Bernu2001,momoi2012}. In the bilayer system, however, exchanges involving 4 or more particles are suppressed with decreasing $d/a$ (see Fig.~\ref{fig:jV} and Appendix \ref{app:large_cyc}), leading to reduced frustration and enhanced stability of the ferromagnetic state to smaller $r_s$. 

\subsubsection{Phase III+IV}

\begin{figure}
\includegraphics[]{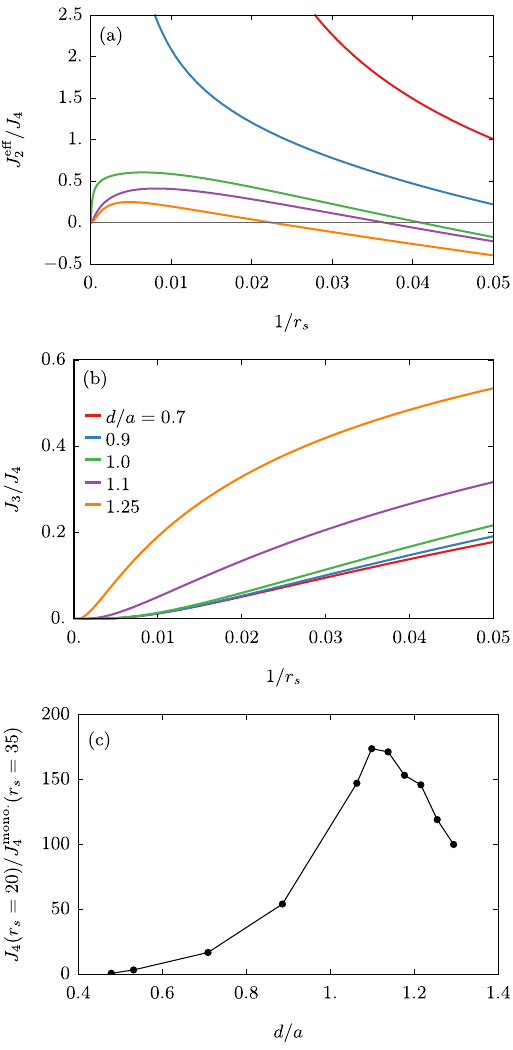}
\caption{Exchange couplings in Phases III and IV. (a) Ratio $J_2^\text{eff}/J_4$ as a function of $1/r_s$ for $d/a = 0.7, 0.9 ,1.0, 1.1, 1.25$, which includes Phase III ($0.46  \lesssim d/a \lesssim 1.1$) and IV ($1.1 \lesssim d/a \lesssim 1.3$). (b) Ratio of 3-particle (which determines the further-neighbor exchange) to the 4-particle exchange. $J_3$ is significantly suppressed with decreasing $d/a$.  (c) $J_4$ as a function of $d/a$ at $r_s=20$, roughly the smallest $r_s$ down to which the bilayer WC is estimated to be stable \cite{rapisarda1998}, normalized by the monolayer value $J_4^\text{mono.}$ evaluated at the monolayer melting point $r_s=35$.
}
\label{fig:jIII+IV}
\end{figure}

Phases III and IV are connected continuously, and we therefore discuss the effective exchange Hamiltonian in both phases together. (We could also include Phase II in this discussion, but prefer to leave it for next subsection.) Keeping again only the 2, 3, and 4-particle exchanges, the exchange Hamiltonian in Phases III and IV may be written 
    \be
    H_\text{ex}^{\text{(III/IV)}} =  J_2^\text{eff}\sum_{\includegraphics[width=0.025\textwidth]{nn.pdf}} P_2 + J_4 \sum_{\includegraphics[width=0.025\textwidth]{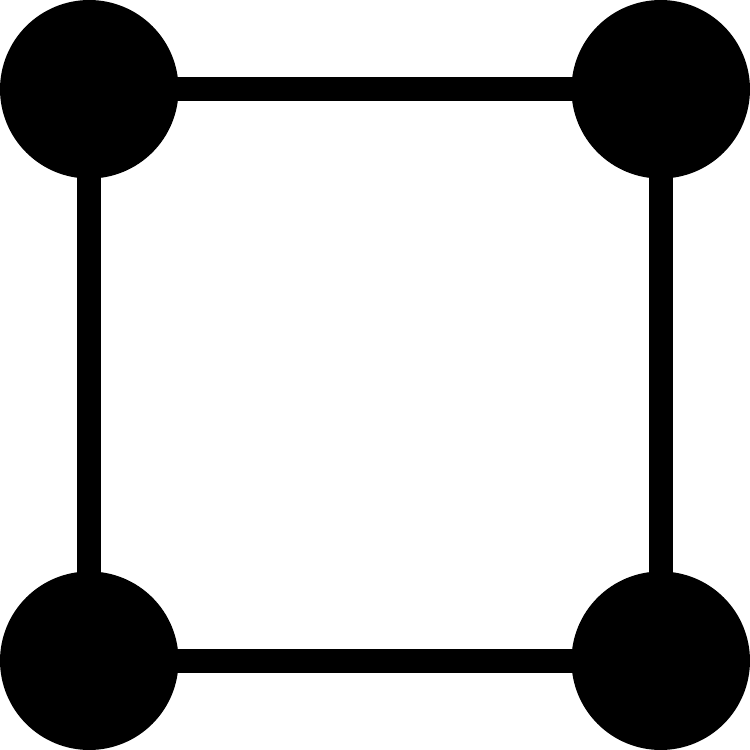}} (P_4 + P_4^{-1}) + \ldots .
    \label{eq:HexIV+III}
    \ee
It is understood that in Phase IV the 4-particle exchange term is a sum over the elementary rhombus, which becomes a square in Phase III. As in the discussion of Phase V, we have decomposed the 3-particle exchange in terms of 2-particle exchanges. This leads to an effective nearest-neighbor 2-particle exchange $J_2^\text{eff} = J_2 - 2(J_3 + \tilde J_3)$, as well as a contribution to further-neighbor exchange, indicated by the ellipses. Our results for the ratio $J_2^\text{eff}/J_4$ are presented in Fig.~\ref{fig:jIII+IV}a. The non-trivial evolution of $J_2^\text{eff}/J_4$ with $d/a$ and $r_s$ has interesting implications for the possible magnetic phases, as will be described below. In Fig.~\ref{fig:jIII+IV}b we show  the ratio $J_3/J_4$. We find $J_3$ is suppressed with decreasing $d/a$, which implies a diminishing importance of the further-neighbor exchange.

An ED study of the Hamiltonian \eqref{eq:HexIV+III} without the further-neighbor 2-particle exchange terms was carried out in Ref.~\cite{lauchli2005}. The following evolution was found: For $J_2^\text{eff}/J_4 \ll 1$ the ground state is the 4-sublattice antiferromagnet mentioned in Sec.~\ref{sec:sIV} and ~\ref{sec:sIII}. Upon increasing $J_2^\text{eff}/J_4$ there is a transition at $J_2^\text{eff}/J_4 \approx 0.2$ to a spin-nematic, followed by a transition to a staggered valence-bond solid (VBS) state when $J_2^\text{eff}/J_4 \approx 0.5$. Finally, when $J_2^\text{eff}/J_4 > 1$ there is a transition to a Néel antiferromagnet (Sec.~\ref{sec:sIII}), though we are not aware of any estimates of the transition point. 

Our estimates of the exchanges, shown in Fig.~\ref{fig:jIII+IV}, suggest all the magnetic phases mentioned above may be realized in the BLWC with varying $d/a$ and $r_s$. More accurate estimates of the exchange couplings utilizing, e.g., path-integral Monte Carlo methods \cite{Bernu2001}, as well as an extension of the ED study in \cite{lauchli2005} to include further-neighbor exchange -- which is more important for larger values of $d/a$ (see Fig.~\ref{fig:jIII+IV}b) -- would be useful to clarify this picture.

In Fig.~\ref{fig:jIII+IV}c we compare the magnitude of $J_4$ at $r_s=20$, which is roughly the smallest value of $r_s$ down to which the BLWC is estimated to be stable \cite{rapisarda1998}, to the monolayer $J_4^\text{mono.}$ at the monolayer melting point $r_s \approx 35$ \cite{azadi2024}. The enhanced stability of the BLWC to higher density suggests the magnetic energy scales may be one to two orders of magnitude larger than in the monolayer. 

\subsubsection{Phase II}

\begin{figure}
\includegraphics[]{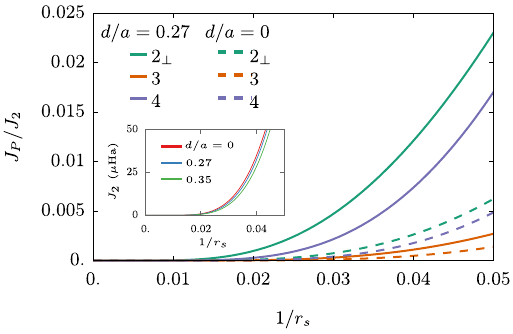}
\caption{Exchange couplings in Phase II of the bilayer WC, normalized by $J_2$. See Fig.~\ref{fig:path_II} for labeling of exchange paths. Couplings are shown for two values of interlayer distance $d/a=0.27$ and $d/a=0$. Inset shows magnitude of $J_2$ for $d/a=0, 0.27, 0.35$.}
\label{fig:jII}
\end{figure}

In Phase II the dominant exchange is $J_2$, the magnetic interactions becoming increasingly anisotropic as $d/a \to 0$; see Fig.~\ref{fig:jII}. For instance, when $d/a \approx 0.27$ and $r_s = 30$, we find $J_{2_\perp}$ (Fig.~\ref{fig:path_II}a) is approximately 200 times smaller than $J_2$. The magnetism of the BLWC in Phase II will therefore be that of a quasi-1D Heisenberg antiferromagnet over the entire relevant range of $r_s$. A similar conclusion regarding the magnetism of the BLWC as $d/a\to 0$ was reached earlier in Ref.~\cite{hirashima2001}. Except in the immediate vicinity of the transition between Phases II and III, the magnitude of $J_2$ is only weakly dependent on the interlayer distance $d/a$ and is of the order of one micro-Hartree near the melting point at $r_s \approx \sqrt 2 \times 35 \approx 50$ (inset of Fig.~\ref{fig:jII}). 

While interlayer tunneling has been neglected throughout this study, as $d/a \to 0$ it is of interest speculate on the effects of weak tunneling with an associated antiferromagnetic 2-particle exchange $J_t$ between nearest-neighbors in the two layers. The strong anisotropy $J_{2_\perp} \ll J_2$ suggests two possible regimes: (i) $J_t \ll J_{2_\perp} \ll J_2$ and (ii) $J_{2_\perp} \ll J_t \ll J_2$. In Case (i) the problem maps to an anisotropic square lattice Heisenberg model in the limit where next-nearest-neighbor exchange dominates over nearest-neighbor exchange. In this case, we may expect a collinear antiferromagnetic state to be stabilized between the layers via  an order-by-disorder mechanism \cite{chandra1990}. In Case (ii) the problem maps to a Heisenberg model on an anisotropic triangular lattice, where the coupling along the chains ($J_2$) dominates the interchain coupling ($J_t$). While there have been a variety of numerical and analytical studies of this regime -- with proposed ground states including either 1D or 2D spin-liquids, as well as collinear antiferromagnetic order \cite{[][{, and references therein}]Starykh2015} -- the nature of the ground state remains, to the best of our knowledge, an open question. Though it is interesting in principle, it seems unlikely current BLWC systems could shed light on the issue given the exceedingly low energy scales associated with both $J_{2_\perp}$ and $J_t$.

\section{Discussion}

\begin{figure}[]
\includegraphics[width=\columnwidth]{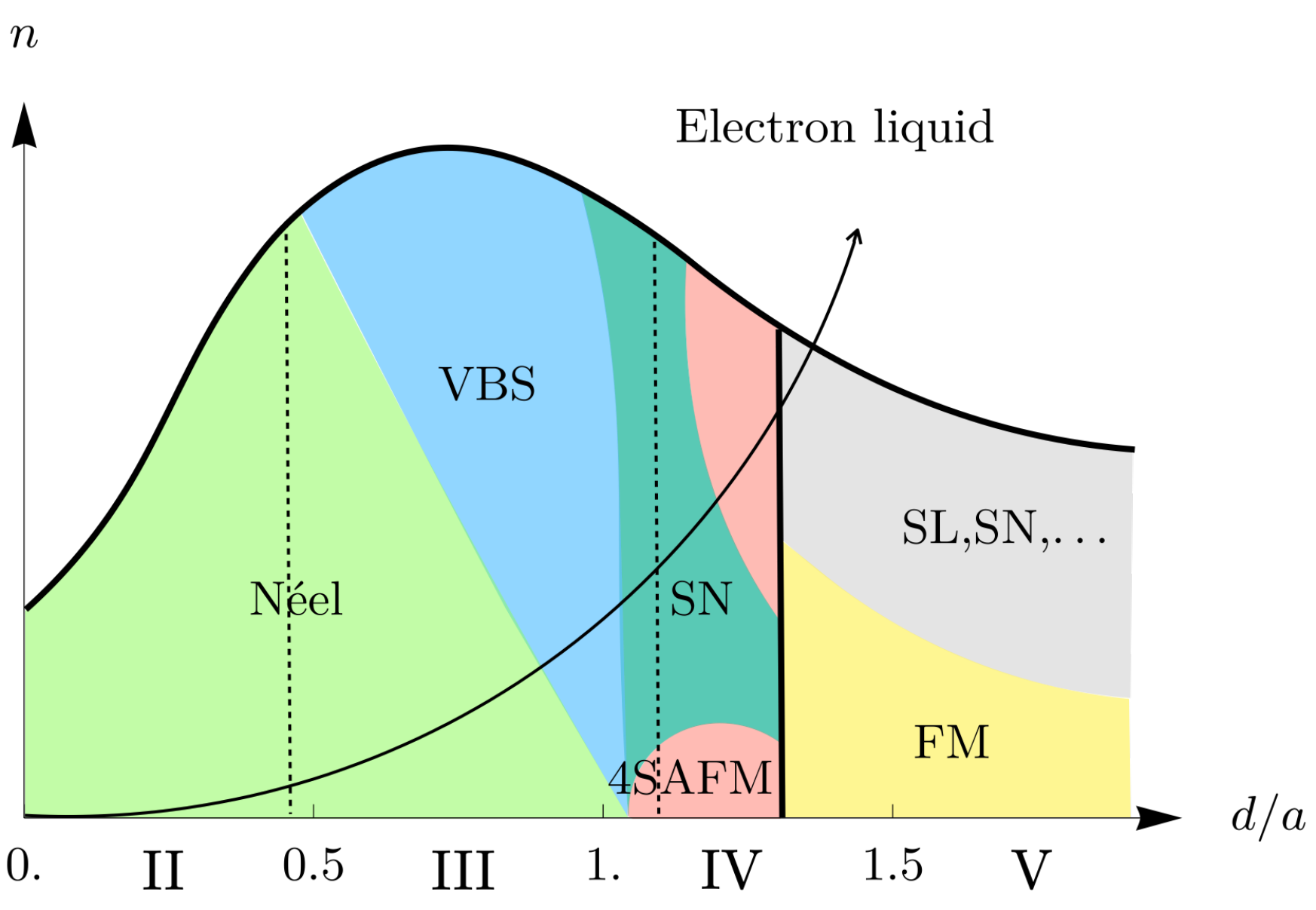}
\caption{\label{fig:phase_diag} Schematic zero-temperature phase diagram of the bilayer 2DEG as a function of $d/a$ and density $n$. Roman numerals indicate the different structural phases. The various magnetic phases are ferromagnet (FM), 4-sublattic antiferromagnet (4SAFM), Néel, spin-nematic (SN), valence-bond solid (VBS), and spin-liquid (SL). At higher density in Phase V we have included some of the candidate magnetic ground states of the multiple-spin exchange model on the triangular lattice, the ellipses indicating that the true ground state depends sensitively on larger ring exchanges not considered here. Arrowed curve indicates a possible experimental path taken by varying $n$ at fixed interlayer separation $d$. The phase boundaries are sketched, being motivated by earlier work on the liquid-solid transition \cite{rapisarda1996} the present estimates of the exchange couplings in the different geometries. More accurate estimates of the exchanges at finite $r_s$, as well as the effects of next-nearest-neighbor exchange in Phases II and III, are likely to alter details of the magnetic phase diagram near the melting transition. Effects of finite $r_s$ on the boundaries between structural phases have also not been taken into account.}
\end{figure}

We have analyzed the multiple spin-exchange processes in the BLWC, from which the magnetic ordering of the crystal is inferred. In the extreme dilute limit $r_s \to \infty$, we have obtained asymptotically exact results for the magnetic phases of the BLWC as a function of the interlayer distance $d/a$ across the four different structural phases, finding both ferromagnetic and multi-sublattice antiferromagnetic orders (see Fig.~\ref{fig:phase_diag_1d}). These results have also been extended to finite $r_s$ -- where the semiclassical approach is only approximate -- and we have argued for a rich set of possible magnetic phases, including a spin-nematic phase, a VBS phase, and possible spin liquids \footnote{The putative transition between the VBS and Néel state in Phase III is of particular interest, as such a transition has been argued to proceed via a deconfined quantum critical point. While interesting field-theories for the staggered VBS to Néel transition have been proposed \cite{Xu2011}, numerical work \cite{Sen2010} suggests the transition in the staggered VBS case is likely to be first-order.}. Estimates of the exchanges suggest significantly enhanced magnetic energy scales at intermediate $d/a$ as compared to the monolayer system. 

Our findings are summarized in the schematic phase diagram in Fig.~\ref{fig:phase_diag}. The magnetic phase boundaries are qualitative sketches motivated by our estimates of the exchange couplings. We caution that certain features may be artifacts of the semiclassical approximation and the detailed structure of the phase diagram at finite $r_s$ will require more accurate estimates of the exchanges and analysis of the associated magnetic Hamiltonians. In experimental bilayer systems one typically works with a fixed interlayer distance, sweeping the carrier density in the two layers -- such a cut through the phase diagram is shown in Fig.~\ref{fig:phase_diag}. 

Given the recent discovery of BLWC phases in a bilayer MoSe$_2$ system \cite{Zhou2021}, it is of interest to estimate the magnetic energy scales associated to the phases predicted here. In MoSe$_2$ the effective Hartree energy is roughly 0.7 eV. Depending on the interlayer distance and the density, this yields magnetic energy scales on the order of a few mK up to several hundred mK. The upper end of this range, which is within current experimental reach \cite{sung2023}, occurs in the vicinity of Phase III where one may hope to observe a possible VBS or spin-nematic.

There are a number of natural extensions of the present work. For instance, in the case of bilayers with unequal density there are yet more crystal geometries that can be realized \cite{Vilk1985,Zhou2021}, likely with further interesting or exotic magnetic phases. The effects of quenched disorder -- present in any real 2DEG system -- on the magnetic phases we have discussed may also be analyzed with semiclassical methods \cite{voelker2001}.

Multiple-spin exchange is not the only mechanism for magnetic interactions in the WC. Another interesting possibility is kinetic magnetism \cite{kim2022,kim2024}\footnote{The importance of kinetic magnetism in triangular lattice Hubbard models has also been recently emphasized \cite{ciorciaro2023,*morera2024}.}, where magnetic couplings result from the quantum dynamics of defects in the crystal. Such effects are especially important if the WC contains a finite concentration of ground state defects \cite{kim2024}, if defects are generated due to quenched disorder \cite{kim2022}, or when the BLWC is doped away from commensurate density ratios between layers. The defect hopping amplitudes may be estimated with the same semiclassical methods used here. Interestingly, in the case of the monolayer WC, the energy scales associated with kinetic magnetism are significantly larger than those associated with ring exchange \cite{kim2022,kim2024}. 

The ring-exchange magnetism we have described is due to local processes in the BLWC. Because strong local crystalline correlations are built into the \textit{liquid phase} of the 2DEG at large $r_s$, one may expect that similar processes are relevant \cite{castaing1980}. The magnetic characteristics of the bilayer liquid ``inherited" from the proximate crystal may be quite interesting. 

\begin{figure}
\includegraphics[]{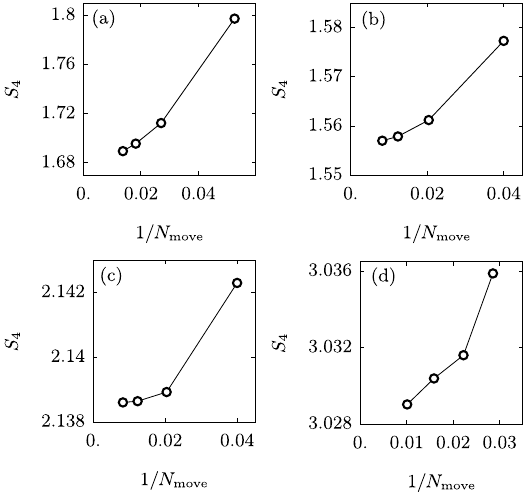}
\caption{Convergence of the 4-particle exchange action $S_4$ with $N_\text{move}$ for representative values of $d/a$ in each BLWC geometry. (a) $d/a \approx 1.5$ (Phase V), (b) $d/a \approx 1.2$ (Phase IV), (c) $d/a \approx 0.71$ (Phase III), (d) $d/a \approx 0.27$ (Phase II). The action integral is discretized into $M=16$ slices.}
\label{fig:sconv}
\end{figure}

\begin{figure}
\includegraphics[]{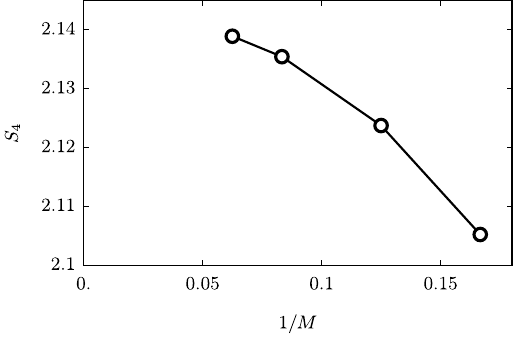}
\caption{Convergence of the 4-particle exchange action $S_4$ with $M$ for $d/a\approx 0.71$ (Phase III). Data is for $N_\text{move}=49$.}
\label{fig:sconvM}
\end{figure}

\begin{acknowledgments}

The authors are grateful to K. S. Kim for many useful discussions, especially concerning technical aspects of the calculations. One of us (I.E.) acknowledges H. Park, J. Sung, P. Volkov, Y. Yang, S. Zhang, and Y. Zhou for helpful discussions on related work, and would also especially like to thank E. Demler and Y. Wang for many fruitful discussions regarding bilayer Wigner crystals. This research project was financially supported by the National Science Foundation Grant No. DMR-2203411 (D.Z. and Z.Z.) and H. I. Romnes Faculty Fellowship provided by the University of Wisconsin-Madison Office of the Vice Chancellor for Research and Graduate Education with funding from the Wisconsin Alumni Research Foundation (A.L.). I.E. was supported by the University of Wisconsin - Madison.
This work was performed in part at Aspen Center for Physics, during the program ``Quantum Matter Through the Lens of Moir\'e Materials", which is supported by National Science Foundation grant PHY-2210452. 
\end{acknowledgments}

\appendix

\section{Convergence}
\label{app:conv}

\begin{figure}
\includegraphics[]{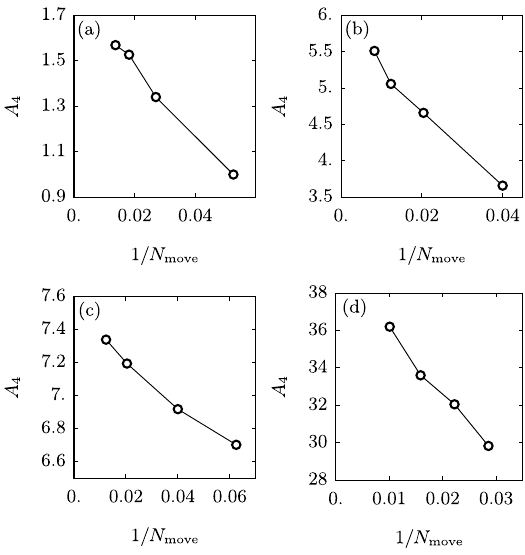}
\caption{Convergence of the 4-particle exchange prefactor $A_4$ with $N_\text{move}$ for representative values of $d/a$ in each BLWC geometry. (a) $d/a \approx 1.5$ (Phase V), (b) $d/a \approx 1.2$ (Phase IV), (c) $d/a \approx 0.71$ (Phase III), (d) $d/a \approx 0.27$ (Phase II). The action integral is discretized into $M=16$ slices.}
\label{fig:aconv}
\end{figure}

To calculate the classical imaginary-time trajectories for a particular ring exchange process we minimize the action \eqref{eq:SP}. This involves two approximations: (1) We allow only a finite number $N_\text{move}$ of electrons to adjust their positions in each layer and (2) the integral \eqref{eq:SP} is discretized into $M$ steps. 

As an example, in Fig.~\ref{fig:sconv} we present data for convergence of $S_4$ with increasing $N_\text{move}$ and $M=16$, for a representative value of $d/a$ in each of the different BLWC geometries. The action is well-converged for $N_\text{move} \gtrsim 100$, the difference between the value of $S_4$ extrapolated to $N\to\infty$ and that at the largest value of $N_\text{move}$ being less than 1\%. The same is true for other values of $d/a$ and other ring exchange processes. 

The convergence of the action with $M$ is shown in Fig.~\ref{fig:sconvM}, taking $S_4$ with $d/a \approx 0.71$ (Phase III) and $N_\text{move}=49$ as a representative example. We find the difference between the value of $S_4$ extrapolated to $M\to \infty$ and that $M=16$ is $\lesssim 0.1\%$. Taken together with the convergence with $N_\text{move}$, we believe our results for the actions should be accurate to within less than 1\%. 

In Fig.~\ref{fig:aconv} we show the convergence of the prefactor $A_4$ at $M=16$ with increasing $N_\text{move}$. The prefactor evidently converges more slowly than the action and, in many cases, the evolution with $N_\text{move}$ does not admit a well-defined extrapolation to $N_\text{move} \to \infty$. In the main text we therefore report the value of $A_P$ for the largest value of $N_\text{move}$ used for a particular ring exchange and value of $d/a$. In those cases where an extrapolation to $N_\text{move} \to \infty$ is possible, we find our errors are $\lesssim 10\%$.

\begin{figure}[t!]
\includegraphics[]{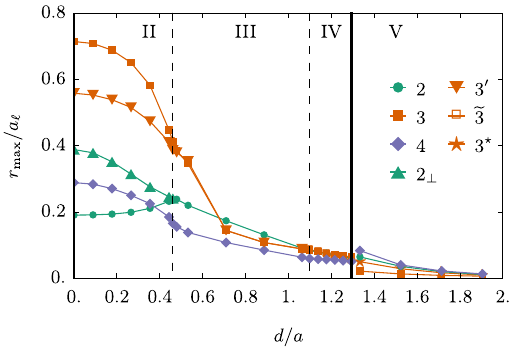}
\caption{Maximal displacement $r_\text{max}$ of a particle away from its equilibrium lattice position in the passive layer as a function of $d/a$ for various exchanges, expressed in units of the lattice constant $a_\ell$ for the corresponding geometry. Labeling corresponds to that in Figs.~\ref{fig:path_V}-\ref{fig:path_II}.}
\label{fig:rmax}
\end{figure}

\begin{figure*}[t!]
\includegraphics[width=\textwidth]{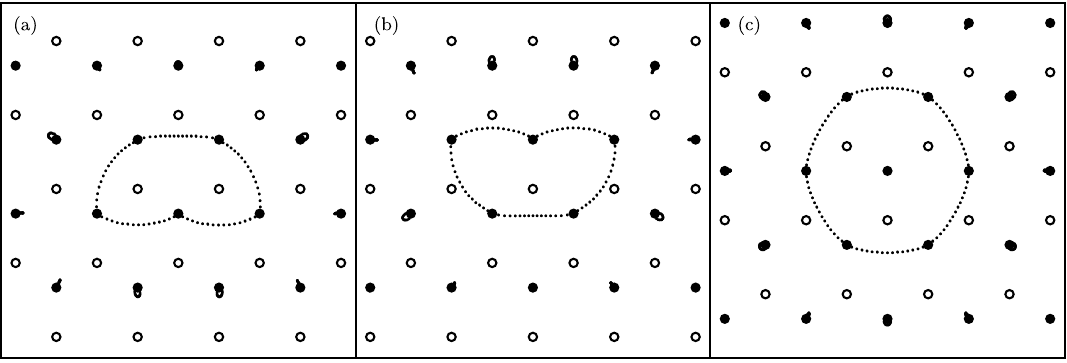}
\caption{Paths for (a,b) 5, and (c) 6-particle exchange in the staggered triangular geometry (Phase V) for interlayer separation $d/a \approx 1.5$. We refer to the inequivalent 5-particle exchange paths in Panels (a) and (b) as 5 and 5$^\star$, respectively.}
\label{fig:large_V}
\end{figure*}

\begin{figure}
\includegraphics[]{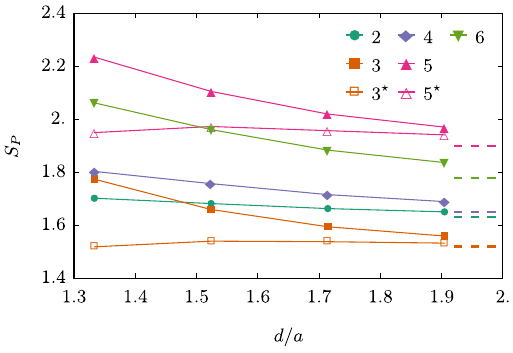}
\caption{Actions for 2,3,4,5, and 6 particle exchange in Phase V. The distinction between paths $3/3^\star$ and $5/5^\star$ is in the nearest neighbors from the opposite layer (see Fig.~\ref{fig:path_V} and Fig.~\ref{fig:large_V}, respectively). Dashed lines are the values of the corresponding actions in the monolayer limit $d\to \infty$.}
\label{fig:sV}
\end{figure}

\section{Role of the passive layer}
\label{app:pass}

When the interlayer coupling is weak ($d/a \gg 1$), electrons in the passive layer do not adjust their positions significantly throughout the exchange process, serving primarily as an ``external" potential to stabilize the crystal in the active layer. In the strong interalyer coupling limit ($d/a \ll 1$), the electrons in the passive layer participate more in the exchange, making larger excursions from their equilibrium positions. 

This evolution is quantified in Fig.~\ref{fig:rmax}, where we present the maximal displacement $r_\text{max}$ of an electron in the passive layer from its equilibrium position as a function of $d/a$. For $d/a \gtrsim 1$, we indeed find that $r_\text{max}$ does not exceed roughly 10\% of the BLWC lattice constant $a_\ell$ for all the exchange process considered. On the other hand, when $d/a \to 0$ the displacements can be on the order of the lattice constant, depending on the particular exchange process. This evolution of the passive layer displacements can also be qualitatively inferred from the paths shown in Figs.~\ref{fig:path_V}-\ref{fig:path_II}.

\section{Larger exchange cycles}
\label{app:large_cyc}

In general, the interlayer coupling tends to reduce the importance of large exchange cycles in the BLWC relative to the monolayer. To the extent we expect such cycles to play a role in determining the magnetic ground state, this will most likely be for larger values of $d/a$. In Fig.~\ref{fig:large_V} we therefore also report the exchange paths associated with 5 and 6-particles exchange in Phase V for $d/a\approx 1.5$, with the corresponding actions shown in Fig.~\ref{fig:sV} as a function $d/a$. Except paths $3^\star$ and $5^\star$ (shown in Figs.~\ref{fig:path_V}c and \ref{fig:large_V}b, respectively), the actions are monotonically increasing with decreasing $d/a$.

\bibliography{bilayer_magnetism}

\begin{thebibliography}{48}%
\makeatletter
\providecommand \@ifxundefined [1]{%
 \@ifx{#1\undefined}
}%
\providecommand \@ifnum [1]{%
 \ifnum #1\expandafter \@firstoftwo
 \else \expandafter \@secondoftwo
 \fi
}%
\providecommand \@ifx [1]{%
 \ifx #1\expandafter \@firstoftwo
 \else \expandafter \@secondoftwo
 \fi
}%
\providecommand \natexlab [1]{#1}%
\providecommand \enquote  [1]{``#1''}%
\providecommand \bibnamefont  [1]{#1}%
\providecommand \bibfnamefont [1]{#1}%
\providecommand \citenamefont [1]{#1}%
\providecommand \href@noop [0]{\@secondoftwo}%
\providecommand \href [0]{\begingroup \@sanitize@url \@href}%
\providecommand \@href[1]{\@@startlink{#1}\@@href}%
\providecommand \@@href[1]{\endgroup#1\@@endlink}%
\providecommand \@sanitize@url [0]{\catcode `\\12\catcode `\$12\catcode
  `\&12\catcode `\#12\catcode `\^12\catcode `\_12\catcode `\%12\relax}%
\providecommand \@@startlink[1]{}%
\providecommand \@@endlink[0]{}%
\providecommand \url  [0]{\begingroup\@sanitize@url \@url }%
\providecommand \@url [1]{\endgroup\@href {#1}{\urlprefix }}%
\providecommand \urlprefix  [0]{URL }%
\providecommand \Eprint [0]{\href }%
\providecommand \doibase [0]{https://doi.org/}%
\providecommand \selectlanguage [0]{\@gobble}%
\providecommand \bibinfo  [0]{\@secondoftwo}%
\providecommand \bibfield  [0]{\@secondoftwo}%
\providecommand \translation [1]{[#1]}%
\providecommand \BibitemOpen [0]{}%
\providecommand \bibitemStop [0]{}%
\providecommand \bibitemNoStop [0]{.\EOS\space}%
\providecommand \EOS [0]{\spacefactor3000\relax}%
\providecommand \BibitemShut  [1]{\csname bibitem#1\endcsname}%
\let\auto@bib@innerbib\@empty
\bibitem [{\citenamefont {Wigner}(1934)}]{wigner1934}%
  \BibitemOpen
  \bibfield  {author} {\bibinfo {author} {\bibfnamefont {E.}~\bibnamefont
  {Wigner}},\ }\bibfield  {title} {\bibinfo {title} {On the interaction of
  electrons in metals},\ }\href {https://doi.org/10.1103/PhysRev.46.1002}
  {\bibfield  {journal} {\bibinfo  {journal} {Phys. Rev.}\ }\textbf {\bibinfo
  {volume} {46}},\ \bibinfo {pages} {1002} (\bibinfo {year}
  {1934})}\BibitemShut {NoStop}%
\bibitem [{\citenamefont {Bonsall}\ and\ \citenamefont
  {Maradudin}(1977)}]{bonsall1977}%
  \BibitemOpen
  \bibfield  {author} {\bibinfo {author} {\bibfnamefont {L.}~\bibnamefont
  {Bonsall}}\ and\ \bibinfo {author} {\bibfnamefont {A.~A.}\ \bibnamefont
  {Maradudin}},\ }\bibfield  {title} {\bibinfo {title} {Some static and
  dynamical properties of a two-dimensional wigner crystal},\ }\href
  {https://doi.org/10.1103/PhysRevB.15.1959} {\bibfield  {journal} {\bibinfo
  {journal} {Phys. Rev. B}\ }\textbf {\bibinfo {volume} {15}},\ \bibinfo
  {pages} {1959} (\bibinfo {year} {1977})}\BibitemShut {NoStop}%
\bibitem [{\citenamefont {Vil'k}\ and\ \citenamefont
  {Monarkha}(1984)}]{Vilk1984}%
  \BibitemOpen
  \bibfield  {author} {\bibinfo {author} {\bibfnamefont {Y.~M.}\ \bibnamefont
  {Vil'k}}\ and\ \bibinfo {author} {\bibfnamefont {Y.~P.}\ \bibnamefont
  {Monarkha}},\ }\href@noop {} {\bibfield  {journal} {\bibinfo  {journal} {Fiz.
  Nizk. Temp.}\ }\textbf {\bibinfo {volume} {10}},\ \bibinfo {pages} {886}
  (\bibinfo {year} {1984})}\BibitemShut {NoStop}%
\bibitem [{\citenamefont {Vil'k}\ and\ \citenamefont
  {Monarkha}(1985)}]{Vilk1985}%
  \BibitemOpen
  \bibfield  {author} {\bibinfo {author} {\bibfnamefont {Y.~M.}\ \bibnamefont
  {Vil'k}}\ and\ \bibinfo {author} {\bibfnamefont {Y.~P.}\ \bibnamefont
  {Monarkha}},\ }\href@noop {} {\bibfield  {journal} {\bibinfo  {journal} {Fiz.
  Nizk. Temp.}\ }\textbf {\bibinfo {volume} {11}},\ \bibinfo {pages} {971}
  (\bibinfo {year} {1985})}\BibitemShut {NoStop}%
\bibitem [{\citenamefont {Falko}(1994)}]{Falko1994}%
  \BibitemOpen
  \bibfield  {author} {\bibinfo {author} {\bibfnamefont {V.~I.}\ \bibnamefont
  {Falko}},\ }\bibfield  {title} {\bibinfo {title} {Optical branch of
  magnetophonons in a double-layer wigner crystal},\ }\href
  {https://doi.org/10.1103/PhysRevB.49.7774} {\bibfield  {journal} {\bibinfo
  {journal} {Phys. Rev. B}\ }\textbf {\bibinfo {volume} {49}},\ \bibinfo
  {pages} {7774} (\bibinfo {year} {1994})}\BibitemShut {NoStop}%
\bibitem [{\citenamefont {Narasimhan}\ and\ \citenamefont
  {Ho}(1995)}]{Narsimhan1995}%
  \BibitemOpen
  \bibfield  {author} {\bibinfo {author} {\bibfnamefont {S.}~\bibnamefont
  {Narasimhan}}\ and\ \bibinfo {author} {\bibfnamefont {T.-L.}\ \bibnamefont
  {Ho}},\ }\bibfield  {title} {\bibinfo {title} {Wigner-crystal phases in
  bilayer quantum hall systems},\ }\href
  {https://doi.org/10.1103/PhysRevB.52.12291} {\bibfield  {journal} {\bibinfo
  {journal} {Phys. Rev. B}\ }\textbf {\bibinfo {volume} {52}},\ \bibinfo
  {pages} {12291} (\bibinfo {year} {1995})}\BibitemShut {NoStop}%
\bibitem [{\citenamefont {Esfarjani}\ and\ \citenamefont
  {Kawazoe}(1995)}]{Esfarjani1995}%
  \BibitemOpen
  \bibfield  {author} {\bibinfo {author} {\bibfnamefont {K.}~\bibnamefont
  {Esfarjani}}\ and\ \bibinfo {author} {\bibfnamefont {Y.}~\bibnamefont
  {Kawazoe}},\ }\bibfield  {title} {\bibinfo {title} {A bilayer of wigner
  crystal in the harmonic approximation},\ }\href
  {https://doi.org/10.1088/0953-8984/7/36/011} {\bibfield  {journal} {\bibinfo
  {journal} {Journal of Physics: Condensed Matter}\ }\textbf {\bibinfo {volume}
  {7}},\ \bibinfo {pages} {7217} (\bibinfo {year} {1995})}\BibitemShut
  {NoStop}%
\bibitem [{\citenamefont {Goldoni}\ and\ \citenamefont
  {Peeters}(1996)}]{goldoni1996}%
  \BibitemOpen
  \bibfield  {author} {\bibinfo {author} {\bibfnamefont {G.}~\bibnamefont
  {Goldoni}}\ and\ \bibinfo {author} {\bibfnamefont {F.~M.}\ \bibnamefont
  {Peeters}},\ }\bibfield  {title} {\bibinfo {title} {Stability, dynamical
  properties, and melting of a classical bilayer wigner crystal},\ }\href
  {https://doi.org/10.1103/PhysRevB.53.4591} {\bibfield  {journal} {\bibinfo
  {journal} {Phys. Rev. B}\ }\textbf {\bibinfo {volume} {53}},\ \bibinfo
  {pages} {4591} (\bibinfo {year} {1996})}\BibitemShut {NoStop}%
\bibitem [{\citenamefont {Goldoni}\ and\ \citenamefont
  {Peeters}(1997)}]{goldoni1997}%
  \BibitemOpen
  \bibfield  {author} {\bibinfo {author} {\bibfnamefont {G.}~\bibnamefont
  {Goldoni}}\ and\ \bibinfo {author} {\bibfnamefont {F.~M.}\ \bibnamefont
  {Peeters}},\ }\bibfield  {title} {\bibinfo {title} {Wigner crystallization in
  quantum electron bilayers},\ }\href
  {https://doi.org/10.1209/epl/i1997-00544-3} {\bibfield  {journal} {\bibinfo
  {journal} {Europhysics Letters}\ }\textbf {\bibinfo {volume} {37}},\ \bibinfo
  {pages} {293} (\bibinfo {year} {1997})}\BibitemShut {NoStop}%
\bibitem [{Note1()}]{Note1}%
  \BibitemOpen
  \bibinfo {note} {In Ref.~\cite {goldoni1996} it was concluded that there is a
  non-zero range near $d/a=0$ where a single-component triangular lattice is
  stable (Phase I). It was later shown, however, that the single-component
  triangular lattice is stable only at $d/a=0$ \cite {samaj2012}, with the
  aspect ratio in Phase II evolving continuously down to $d/a =
  0$.}\BibitemShut {Stop}%
\bibitem [{\citenamefont {\ifmmode~\acute{S}\else \'{S}\fi{}wierkowski}\ \emph
  {et~al.}(1991)\citenamefont {\ifmmode~\acute{S}\else \'{S}\fi{}wierkowski},
  \citenamefont {Neilson},\ and\ \citenamefont {Szyma\ifmmode~\acute{n}\else
  \'{n}\fi{}ski}}]{swierkowski1991}%
  \BibitemOpen
  \bibfield  {author} {\bibinfo {author} {\bibfnamefont {L.}~\bibnamefont
  {\ifmmode~\acute{S}\else \'{S}\fi{}wierkowski}}, \bibinfo {author}
  {\bibfnamefont {D.}~\bibnamefont {Neilson}},\ and\ \bibinfo {author}
  {\bibfnamefont {J.}~\bibnamefont {Szyma\ifmmode~\acute{n}\else
  \'{n}\fi{}ski}},\ }\bibfield  {title} {\bibinfo {title} {Enhancement of
  wigner crystallization in multiple-quantum-well structures},\ }\href
  {https://doi.org/10.1103/PhysRevLett.67.240} {\bibfield  {journal} {\bibinfo
  {journal} {Phys. Rev. Lett.}\ }\textbf {\bibinfo {volume} {67}},\ \bibinfo
  {pages} {240} (\bibinfo {year} {1991})}\BibitemShut {NoStop}%
\bibitem [{\citenamefont {Rapisarda}\ and\ \citenamefont
  {Senatore}(1996)}]{rapisarda1996}%
  \BibitemOpen
  \bibfield  {author} {\bibinfo {author} {\bibfnamefont {F.}~\bibnamefont
  {Rapisarda}}\ and\ \bibinfo {author} {\bibfnamefont {G.}~\bibnamefont
  {Senatore}},\ }\bibfield  {title} {\bibinfo {title} {Diffusion monte carlo
  study of electrons in two-dimensional layers},\ }\href@noop {} {\bibfield
  {journal} {\bibinfo  {journal} {Australian journal of physics}\ }\textbf
  {\bibinfo {volume} {49}},\ \bibinfo {pages} {161} (\bibinfo {year}
  {1996})}\BibitemShut {NoStop}%
\bibitem [{\citenamefont {Rapisarda}\ and\ \citenamefont
  {Senatore}(1998)}]{rapisarda1998}%
  \BibitemOpen
  \bibfield  {author} {\bibinfo {author} {\bibfnamefont {F.}~\bibnamefont
  {Rapisarda}}\ and\ \bibinfo {author} {\bibfnamefont {G.}~\bibnamefont
  {Senatore}},\ }\bibinfo {title} {Recent progress on the phase diagram of
  coupled electron layers in zero magnetic field},\ in\ \href
  {https://doi.org/10.1007/0-306-47086-1_96} {\emph {\bibinfo {booktitle}
  {Strongly Coupled Coulomb Systems}}},\ \bibinfo {editor} {edited by\ \bibinfo
  {editor} {\bibfnamefont {G.~J.}\ \bibnamefont {Kalman}}, \bibinfo {editor}
  {\bibfnamefont {J.~M.}\ \bibnamefont {Rommel}},\ and\ \bibinfo {editor}
  {\bibfnamefont {K.}~\bibnamefont {Blagoev}}}\ (\bibinfo  {publisher}
  {Springer US},\ \bibinfo {address} {Boston, MA},\ \bibinfo {year} {1998})\
  pp.\ \bibinfo {pages} {529--532}\BibitemShut {NoStop}%
\bibitem [{\citenamefont {Tanatar}\ and\ \citenamefont
  {Ceperley}(1989)}]{Tanatar1989}%
  \BibitemOpen
  \bibfield  {author} {\bibinfo {author} {\bibfnamefont {B.}~\bibnamefont
  {Tanatar}}\ and\ \bibinfo {author} {\bibfnamefont {D.~M.}\ \bibnamefont
  {Ceperley}},\ }\bibfield  {title} {\bibinfo {title} {Ground state of the
  two-dimensional electron gas},\ }\href
  {https://doi.org/10.1103/PhysRevB.39.5005} {\bibfield  {journal} {\bibinfo
  {journal} {Phys. Rev. B}\ }\textbf {\bibinfo {volume} {39}},\ \bibinfo
  {pages} {5005} (\bibinfo {year} {1989})}\BibitemShut {NoStop}%
\bibitem [{\citenamefont {Attaccalite}\ \emph {et~al.}(2002)\citenamefont
  {Attaccalite}, \citenamefont {Moroni}, \citenamefont {Gori-Giorgi},\ and\
  \citenamefont {Bachelet}}]{Attaccalite2002}%
  \BibitemOpen
  \bibfield  {author} {\bibinfo {author} {\bibfnamefont {C.}~\bibnamefont
  {Attaccalite}}, \bibinfo {author} {\bibfnamefont {S.}~\bibnamefont {Moroni}},
  \bibinfo {author} {\bibfnamefont {P.}~\bibnamefont {Gori-Giorgi}},\ and\
  \bibinfo {author} {\bibfnamefont {G.~B.}\ \bibnamefont {Bachelet}},\
  }\bibfield  {title} {\bibinfo {title} {Correlation energy and spin
  polarization in the 2d electron gas},\ }\href
  {https://doi.org/10.1103/PhysRevLett.88.256601} {\bibfield  {journal}
  {\bibinfo  {journal} {Phys. Rev. Lett.}\ }\textbf {\bibinfo {volume} {88}},\
  \bibinfo {pages} {256601} (\bibinfo {year} {2002})}\BibitemShut {NoStop}%
\bibitem [{\citenamefont {Drummond}\ and\ \citenamefont
  {Needs}(2009)}]{Drummond2009}%
  \BibitemOpen
  \bibfield  {author} {\bibinfo {author} {\bibfnamefont {N.~D.}\ \bibnamefont
  {Drummond}}\ and\ \bibinfo {author} {\bibfnamefont {R.~J.}\ \bibnamefont
  {Needs}},\ }\bibfield  {title} {\bibinfo {title} {Phase diagram of the
  low-density two-dimensional homogeneous electron gas},\ }\href
  {https://doi.org/10.1103/PhysRevLett.102.126402} {\bibfield  {journal}
  {\bibinfo  {journal} {Phys. Rev. Lett.}\ }\textbf {\bibinfo {volume} {102}},\
  \bibinfo {pages} {126402} (\bibinfo {year} {2009})}\BibitemShut {NoStop}%
\bibitem [{\citenamefont {Azadi}\ \emph {et~al.}(2024)\citenamefont {Azadi},
  \citenamefont {Drummond},\ and\ \citenamefont {Vinko}}]{azadi2024}%
  \BibitemOpen
  \bibfield  {author} {\bibinfo {author} {\bibfnamefont {S.}~\bibnamefont
  {Azadi}}, \bibinfo {author} {\bibfnamefont {N.~D.}\ \bibnamefont
  {Drummond}},\ and\ \bibinfo {author} {\bibfnamefont {S.~M.}\ \bibnamefont
  {Vinko}},\ }\href {https://arxiv.org/abs/2405.00425} {\bibinfo {title}
  {Quantum monte carlo study of the phase diagram of the two-dimensional
  uniform electron liquid}} (\bibinfo {year} {2024}),\ \Eprint
  {https://arxiv.org/abs/2405.00425} {arXiv:2405.00425 [cond-mat.str-el]}
  \BibitemShut {NoStop}%
\bibitem [{\citenamefont {Zhou}\ \emph {et~al.}(2021)\citenamefont {Zhou},
  \citenamefont {Sung}, \citenamefont {Brutschea}, \citenamefont {Esterlis},
  \citenamefont {Wang}, \citenamefont {Scuri}, \citenamefont {Gelly},
  \citenamefont {Heo}, \citenamefont {Taniguchi}, \citenamefont {Watanabe},
  \citenamefont {Zar{\'{a}}nd}, \citenamefont {Lukin}, \citenamefont {Kim},
  \citenamefont {Demler},\ and\ \citenamefont {Park}}]{Zhou2021}%
  \BibitemOpen
  \bibfield  {author} {\bibinfo {author} {\bibfnamefont {Y.}~\bibnamefont
  {Zhou}}, \bibinfo {author} {\bibfnamefont {J.}~\bibnamefont {Sung}}, \bibinfo
  {author} {\bibfnamefont {E.}~\bibnamefont {Brutschea}}, \bibinfo {author}
  {\bibfnamefont {I.}~\bibnamefont {Esterlis}}, \bibinfo {author}
  {\bibfnamefont {Y.}~\bibnamefont {Wang}}, \bibinfo {author} {\bibfnamefont
  {G.}~\bibnamefont {Scuri}}, \bibinfo {author} {\bibfnamefont {R.~J.}\
  \bibnamefont {Gelly}}, \bibinfo {author} {\bibfnamefont {H.}~\bibnamefont
  {Heo}}, \bibinfo {author} {\bibfnamefont {T.}~\bibnamefont {Taniguchi}},
  \bibinfo {author} {\bibfnamefont {K.}~\bibnamefont {Watanabe}}, \bibinfo
  {author} {\bibfnamefont {G.}~\bibnamefont {Zar{\'{a}}nd}}, \bibinfo {author}
  {\bibfnamefont {M.~D.}\ \bibnamefont {Lukin}}, \bibinfo {author}
  {\bibfnamefont {P.}~\bibnamefont {Kim}}, \bibinfo {author} {\bibfnamefont
  {E.}~\bibnamefont {Demler}},\ and\ \bibinfo {author} {\bibfnamefont
  {H.}~\bibnamefont {Park}},\ }\bibfield  {title} {\bibinfo {title} {{Bilayer
  Wigner crystals in a transition metal dichalcogenide heterostructure}},\
  }\href {https://doi.org/10.1038/s41586-021-03560-w} {\bibfield  {journal}
  {\bibinfo  {journal} {Nature}\ }\textbf {\bibinfo {volume} {595}},\ \bibinfo
  {pages} {48} (\bibinfo {year} {2021})}\BibitemShut {NoStop}%
\bibitem [{\citenamefont {Smole{\'{n}}ski}\ \emph {et~al.}(2021)\citenamefont
  {Smole{\'{n}}ski}, \citenamefont {Dolgirev}, \citenamefont {Kuhlenkamp},
  \citenamefont {Popert}, \citenamefont {Shimazaki}, \citenamefont {Back},
  \citenamefont {Lu}, \citenamefont {Kroner}, \citenamefont {Watanabe},
  \citenamefont {Taniguchi}, \citenamefont {Esterlis}, \citenamefont {Demler},\
  and\ \citenamefont {Imamoğlu}}]{Smolenski2021}%
  \BibitemOpen
  \bibfield  {author} {\bibinfo {author} {\bibfnamefont {T.}~\bibnamefont
  {Smole{\'{n}}ski}}, \bibinfo {author} {\bibfnamefont {P.~E.}\ \bibnamefont
  {Dolgirev}}, \bibinfo {author} {\bibfnamefont {C.}~\bibnamefont
  {Kuhlenkamp}}, \bibinfo {author} {\bibfnamefont {A.}~\bibnamefont {Popert}},
  \bibinfo {author} {\bibfnamefont {Y.}~\bibnamefont {Shimazaki}}, \bibinfo
  {author} {\bibfnamefont {P.}~\bibnamefont {Back}}, \bibinfo {author}
  {\bibfnamefont {X.}~\bibnamefont {Lu}}, \bibinfo {author} {\bibfnamefont
  {M.}~\bibnamefont {Kroner}}, \bibinfo {author} {\bibfnamefont
  {K.}~\bibnamefont {Watanabe}}, \bibinfo {author} {\bibfnamefont
  {T.}~\bibnamefont {Taniguchi}}, \bibinfo {author} {\bibfnamefont
  {I.}~\bibnamefont {Esterlis}}, \bibinfo {author} {\bibfnamefont
  {E.}~\bibnamefont {Demler}},\ and\ \bibinfo {author} {\bibfnamefont
  {A.}~\bibnamefont {Imamoğlu}},\ }\bibfield  {title} {\bibinfo {title}
  {{Signatures of Wigner crystal of electrons in a monolayer semiconductor}},\
  }\href {https://doi.org/10.1038/s41586-021-03590-4} {\bibfield  {journal}
  {\bibinfo  {journal} {Nature}\ }\textbf {\bibinfo {volume} {595}},\ \bibinfo
  {pages} {53} (\bibinfo {year} {2021})}\BibitemShut {NoStop}%
\bibitem [{\citenamefont {Thouless}(1965)}]{thouless1965}%
  \BibitemOpen
  \bibfield  {author} {\bibinfo {author} {\bibfnamefont {D.~J.}\ \bibnamefont
  {Thouless}},\ }\bibfield  {title} {\bibinfo {title} {Exchange in solid 3he
  and the heisenberg hamiltonian},\ }\href
  {https://doi.org/10.1088/0370-1328/86/5/301} {\bibfield  {journal} {\bibinfo
  {journal} {Proceedings of the Physical Society}\ }\textbf {\bibinfo {volume}
  {86}},\ \bibinfo {pages} {893} (\bibinfo {year} {1965})}\BibitemShut
  {NoStop}%
\bibitem [{\citenamefont {Roger}\ \emph {et~al.}(1983)\citenamefont {Roger},
  \citenamefont {Hetherington},\ and\ \citenamefont {Delrieu}}]{roger1983}%
  \BibitemOpen
  \bibfield  {author} {\bibinfo {author} {\bibfnamefont {M.}~\bibnamefont
  {Roger}}, \bibinfo {author} {\bibfnamefont {J.~H.}\ \bibnamefont
  {Hetherington}},\ and\ \bibinfo {author} {\bibfnamefont {J.~M.}\ \bibnamefont
  {Delrieu}},\ }\bibfield  {title} {\bibinfo {title} {Magnetism in solid
  $^{3}\mathrm{He}$},\ }\href {https://doi.org/10.1103/RevModPhys.55.1}
  {\bibfield  {journal} {\bibinfo  {journal} {Rev. Mod. Phys.}\ }\textbf
  {\bibinfo {volume} {55}},\ \bibinfo {pages} {1} (\bibinfo {year}
  {1983})}\BibitemShut {NoStop}%
\bibitem [{\citenamefont {Roger}(1984)}]{roger1984}%
  \BibitemOpen
  \bibfield  {author} {\bibinfo {author} {\bibfnamefont {M.}~\bibnamefont
  {Roger}},\ }\bibfield  {title} {\bibinfo {title} {Multiple exchange in
  $^{3}\mathrm{He}$ and in the wigner solid},\ }\href
  {https://doi.org/10.1103/PhysRevB.30.6432} {\bibfield  {journal} {\bibinfo
  {journal} {Phys. Rev. B}\ }\textbf {\bibinfo {volume} {30}},\ \bibinfo
  {pages} {6432} (\bibinfo {year} {1984})}\BibitemShut {NoStop}%
\bibitem [{\citenamefont {Katano}\ and\ \citenamefont
  {Hirashima}(2000)}]{katano2000}%
  \BibitemOpen
  \bibfield  {author} {\bibinfo {author} {\bibfnamefont {M.}~\bibnamefont
  {Katano}}\ and\ \bibinfo {author} {\bibfnamefont {D.~S.}\ \bibnamefont
  {Hirashima}},\ }\bibfield  {title} {\bibinfo {title} {Multiple-spin exchange
  in a two-dimensional wigner crystal},\ }\href
  {https://doi.org/10.1103/PhysRevB.62.2573} {\bibfield  {journal} {\bibinfo
  {journal} {Phys. Rev. B}\ }\textbf {\bibinfo {volume} {62}},\ \bibinfo
  {pages} {2573} (\bibinfo {year} {2000})}\BibitemShut {NoStop}%
\bibitem [{\citenamefont {Sudip~Chakravarty}\ and\ \citenamefont
  {Voelker}(1999)}]{chakravarty1999}%
  \BibitemOpen
  \bibfield  {author} {\bibinfo {author} {\bibfnamefont {C.~N.}\ \bibnamefont
  {Sudip~Chakravarty}, \bibfnamefont {Steven~Kivelson}}\ and\ \bibinfo {author}
  {\bibfnamefont {K.}~\bibnamefont {Voelker}},\ }\bibfield  {title} {\bibinfo
  {title} {Wigner glass, spin liquids and the metal-insulator transition},\
  }\href {https://doi.org/10.1080/13642819908214845} {\bibfield  {journal}
  {\bibinfo  {journal} {Philosophical Magazine B}\ }\textbf {\bibinfo {volume}
  {79}},\ \bibinfo {pages} {859} (\bibinfo {year} {1999})},\ \Eprint
  {https://arxiv.org/abs/https://doi.org/10.1080/13642819908214845}
  {https://doi.org/10.1080/13642819908214845} \BibitemShut {NoStop}%
\bibitem [{\citenamefont {Voelker}\ and\ \citenamefont
  {Chakravarty}(2001)}]{voelker2001}%
  \BibitemOpen
  \bibfield  {author} {\bibinfo {author} {\bibfnamefont {K.}~\bibnamefont
  {Voelker}}\ and\ \bibinfo {author} {\bibfnamefont {S.}~\bibnamefont
  {Chakravarty}},\ }\bibfield  {title} {\bibinfo {title} {Multiparticle ring
  exchange in the wigner glass and its possible relevance to strongly
  interacting two-dimensional electron systems in the presence of disorder},\
  }\href {https://doi.org/10.1103/PhysRevB.64.235125} {\bibfield  {journal}
  {\bibinfo  {journal} {Phys. Rev. B}\ }\textbf {\bibinfo {volume} {64}},\
  \bibinfo {pages} {235125} (\bibinfo {year} {2001})}\BibitemShut {NoStop}%
\bibitem [{\citenamefont {Coleman}(1979)}]{Coleman1979}%
  \BibitemOpen
  \bibfield  {author} {\bibinfo {author} {\bibfnamefont {S.}~\bibnamefont
  {Coleman}},\ }\bibinfo {title} {The uses of instantons},\ in\ \href
  {https://doi.org/10.1007/978-1-4684-0991-8_16} {\emph {\bibinfo {booktitle}
  {The Whys of Subnuclear Physics}}},\ \bibinfo {editor} {edited by\ \bibinfo
  {editor} {\bibfnamefont {A.}~\bibnamefont {Zichichi}}}\ (\bibinfo
  {publisher} {Springer US},\ \bibinfo {address} {Boston, MA},\ \bibinfo {year}
  {1979})\ pp.\ \bibinfo {pages} {805--941}\BibitemShut {NoStop}%
\bibitem [{\citenamefont {Hirashima}(2001)}]{hirashima2001}%
  \BibitemOpen
  \bibfield  {author} {\bibinfo {author} {\bibfnamefont {D.~S.}\ \bibnamefont
  {Hirashima}},\ }\bibfield  {title} {\bibinfo {title} {Magnetism of a bilayer
  wigner crystal},\ }\href@noop {} {\bibfield  {journal} {\bibinfo  {journal}
  {Journal of the Physical Society of Japan}\ }\textbf {\bibinfo {volume}
  {70}},\ \bibinfo {pages} {931} (\bibinfo {year} {2001})}\BibitemShut
  {NoStop}%
\bibitem [{\citenamefont {Bernu}\ \emph {et~al.}(2001)\citenamefont {Bernu},
  \citenamefont {C\^andido},\ and\ \citenamefont {Ceperley}}]{Bernu2001}%
  \BibitemOpen
  \bibfield  {author} {\bibinfo {author} {\bibfnamefont {B.}~\bibnamefont
  {Bernu}}, \bibinfo {author} {\bibfnamefont {L.}~\bibnamefont {C\^andido}},\
  and\ \bibinfo {author} {\bibfnamefont {D.~M.}\ \bibnamefont {Ceperley}},\
  }\bibfield  {title} {\bibinfo {title} {Exchange frequencies in the 2d wigner
  crystal},\ }\href {https://doi.org/10.1103/PhysRevLett.86.870} {\bibfield
  {journal} {\bibinfo  {journal} {Phys. Rev. Lett.}\ }\textbf {\bibinfo
  {volume} {86}},\ \bibinfo {pages} {870} (\bibinfo {year} {2001})}\BibitemShut
  {NoStop}%
\bibitem [{Note2()}]{Note2}%
  \BibitemOpen
  \bibinfo {note} {Throughout the text we quote approximate, as opposed to
  exact, values of $d/a$, writing $d/a \approx \protect \ldots $. The reason is
  that, in the numerical calculations, we actually fix the ratio $d/a_\ell $,
  where $a_\ell $ is the lattice constant for a given BLWC geometry. The
  interparticle distance $a$ is related to $a_\ell $ according to $a=\protect
  \sqrt {A_c/\pi }$, where $A_c$ is the unit cell area. In the different
  geometries we have $A_c = \protect \sqrt 3 a_\ell ^2/2$ (Phase V), $A_c =
  \protect \qopname \relax o{sin}\theta a_\ell ^2$ (Phase IV), $A_c = a_\ell
  ^2$ (Phase III), and $A_c = r a_\ell ^2$ (Phase II).}\BibitemShut {Stop}%
\bibitem [{\citenamefont {L\"auchli}\ \emph {et~al.}(2005)\citenamefont
  {L\"auchli}, \citenamefont {Domenge}, \citenamefont {Lhuillier},
  \citenamefont {Sindzingre},\ and\ \citenamefont {Troyer}}]{lauchli2005}%
  \BibitemOpen
  \bibfield  {author} {\bibinfo {author} {\bibfnamefont {A.}~\bibnamefont
  {L\"auchli}}, \bibinfo {author} {\bibfnamefont {J.~C.}\ \bibnamefont
  {Domenge}}, \bibinfo {author} {\bibfnamefont {C.}~\bibnamefont {Lhuillier}},
  \bibinfo {author} {\bibfnamefont {P.}~\bibnamefont {Sindzingre}},\ and\
  \bibinfo {author} {\bibfnamefont {M.}~\bibnamefont {Troyer}},\ }\bibfield
  {title} {\bibinfo {title} {Two-step restoration of su(2) symmetry in a
  frustrated ring-exchange magnet},\ }\href
  {https://doi.org/10.1103/PhysRevLett.95.137206} {\bibfield  {journal}
  {\bibinfo  {journal} {Phys. Rev. Lett.}\ }\textbf {\bibinfo {volume} {95}},\
  \bibinfo {pages} {137206} (\bibinfo {year} {2005})}\BibitemShut {NoStop}%
\bibitem [{Note3()}]{Note3}%
  \BibitemOpen
  \bibinfo {note} {There are two symmetry-related 4-particle exchange paths in
  Phase II (Fig.~\ref {fig:path_II}d) and the exchange coupling $J_4$ is
  therefore multiplied by a factor of two.}\BibitemShut {Stop}%
\bibitem [{Note4()}]{Note4}%
  \BibitemOpen
  \bibinfo {note} {When $d/a=0$ the paths become identical to the same paths
  for a single-layer WC, although the meaning of 3 and 4 particle exchange is
  different from the single-layer case.}\BibitemShut {Stop}%
\bibitem [{\citenamefont {Kim}\ \emph {et~al.}(2024)\citenamefont {Kim},
  \citenamefont {Esterlis}, \citenamefont {Murthy},\ and\ \citenamefont
  {Kivelson}}]{kim2024}%
  \BibitemOpen
  \bibfield  {author} {\bibinfo {author} {\bibfnamefont {K.-S.}\ \bibnamefont
  {Kim}}, \bibinfo {author} {\bibfnamefont {I.}~\bibnamefont {Esterlis}},
  \bibinfo {author} {\bibfnamefont {C.}~\bibnamefont {Murthy}},\ and\ \bibinfo
  {author} {\bibfnamefont {S.~A.}\ \bibnamefont {Kivelson}},\ }\bibfield
  {title} {\bibinfo {title} {Dynamical defects in a two-dimensional wigner
  crystal: Self-doping and kinetic magnetism},\ }\href
  {https://doi.org/10.1103/PhysRevB.109.235130} {\bibfield  {journal} {\bibinfo
   {journal} {Phys. Rev. B}\ }\textbf {\bibinfo {volume} {109}},\ \bibinfo
  {pages} {235130} (\bibinfo {year} {2024})}\BibitemShut {NoStop}%
\bibitem [{\citenamefont {Misguich}\ \emph {et~al.}(1998)\citenamefont
  {Misguich}, \citenamefont {Bernu}, \citenamefont {Lhuillier},\ and\
  \citenamefont {Waldtmann}}]{misguich1998}%
  \BibitemOpen
  \bibfield  {author} {\bibinfo {author} {\bibfnamefont {G.}~\bibnamefont
  {Misguich}}, \bibinfo {author} {\bibfnamefont {B.}~\bibnamefont {Bernu}},
  \bibinfo {author} {\bibfnamefont {C.}~\bibnamefont {Lhuillier}},\ and\
  \bibinfo {author} {\bibfnamefont {C.}~\bibnamefont {Waldtmann}},\ }\bibfield
  {title} {\bibinfo {title} {Spin liquid in the multiple-spin exchange model on
  the triangular lattice: ${}^{3}\mathrm{He}$ on graphite},\ }\href
  {https://doi.org/10.1103/PhysRevLett.81.1098} {\bibfield  {journal} {\bibinfo
   {journal} {Phys. Rev. Lett.}\ }\textbf {\bibinfo {volume} {81}},\ \bibinfo
  {pages} {1098} (\bibinfo {year} {1998})}\BibitemShut {NoStop}%
\bibitem [{\citenamefont {Misguich}\ \emph {et~al.}(1999)\citenamefont
  {Misguich}, \citenamefont {Lhuillier}, \citenamefont {Bernu},\ and\
  \citenamefont {Waldtmann}}]{misguich1999}%
  \BibitemOpen
  \bibfield  {author} {\bibinfo {author} {\bibfnamefont {G.}~\bibnamefont
  {Misguich}}, \bibinfo {author} {\bibfnamefont {C.}~\bibnamefont {Lhuillier}},
  \bibinfo {author} {\bibfnamefont {B.}~\bibnamefont {Bernu}},\ and\ \bibinfo
  {author} {\bibfnamefont {C.}~\bibnamefont {Waldtmann}},\ }\bibfield  {title}
  {\bibinfo {title} {Spin-liquid phase of the multiple-spin exchange
  hamiltonian on the triangular lattice},\ }\href
  {https://doi.org/10.1103/PhysRevB.60.1064} {\bibfield  {journal} {\bibinfo
  {journal} {Phys. Rev. B}\ }\textbf {\bibinfo {volume} {60}},\ \bibinfo
  {pages} {1064} (\bibinfo {year} {1999})}\BibitemShut {NoStop}%
\bibitem [{\citenamefont {Momoi}\ \emph {et~al.}(2012)\citenamefont {Momoi},
  \citenamefont {Sindzingre},\ and\ \citenamefont {Kubo}}]{momoi2012}%
  \BibitemOpen
  \bibfield  {author} {\bibinfo {author} {\bibfnamefont {T.}~\bibnamefont
  {Momoi}}, \bibinfo {author} {\bibfnamefont {P.}~\bibnamefont {Sindzingre}},\
  and\ \bibinfo {author} {\bibfnamefont {K.}~\bibnamefont {Kubo}},\ }\bibfield
  {title} {\bibinfo {title} {Spin nematic order in multiple-spin exchange
  models on the triangular lattice},\ }\href
  {https://doi.org/10.1103/PhysRevLett.108.057206} {\bibfield  {journal}
  {\bibinfo  {journal} {Phys. Rev. Lett.}\ }\textbf {\bibinfo {volume} {108}},\
  \bibinfo {pages} {057206} (\bibinfo {year} {2012})}\BibitemShut {NoStop}%
\bibitem [{\citenamefont {Chandra}\ \emph {et~al.}(1990)\citenamefont
  {Chandra}, \citenamefont {Coleman},\ and\ \citenamefont
  {Larkin}}]{chandra1990}%
  \BibitemOpen
  \bibfield  {author} {\bibinfo {author} {\bibfnamefont {P.}~\bibnamefont
  {Chandra}}, \bibinfo {author} {\bibfnamefont {P.}~\bibnamefont {Coleman}},\
  and\ \bibinfo {author} {\bibfnamefont {A.~I.}\ \bibnamefont {Larkin}},\
  }\bibfield  {title} {\bibinfo {title} {Ising transition in frustrated
  heisenberg models},\ }\href {https://doi.org/10.1103/PhysRevLett.64.88}
  {\bibfield  {journal} {\bibinfo  {journal} {Phys. Rev. Lett.}\ }\textbf
  {\bibinfo {volume} {64}},\ \bibinfo {pages} {88} (\bibinfo {year}
  {1990})}\BibitemShut {NoStop}%
\bibitem [{\citenamefont {Starykh}(2015)}]{Starykh2015}%
  \BibitemOpen
  \bibfield  {author} {\bibinfo {author} {\bibfnamefont {O.~A.}\ \bibnamefont
  {Starykh}},\ }\bibfield  {title} {\bibinfo {title} {Unusual ordered phases of
  highly frustrated magnets: a review},\ }\href
  {https://doi.org/10.1088/0034-4885/78/5/052502} {\bibfield  {journal}
  {\bibinfo  {journal} {Reports on Progress in Physics}\ }\textbf {\bibinfo
  {volume} {78}},\ \bibinfo {pages} {052502} (\bibinfo {year}
  {2015})}\BibitemShut {NoStop}%
\bibitem [{Note5()}]{Note5}%
  \BibitemOpen
  \bibinfo {note} {The putative transition between the VBS and Néel state in
  Phase III is of particular interest, as such a transition has been argued to
  proceed via a deconfined quantum critical point. While interesting
  field-theories for the staggered VBS to Néel transition have been proposed
  \cite {Xu2011}, numerical work \cite {Sen2010} suggests the transition in the
  staggered VBS case is likely to be first-order.}\BibitemShut {Stop}%
\bibitem [{\citenamefont {Sung}\ \emph {et~al.}(2023)\citenamefont {Sung},
  \citenamefont {Wang}, \citenamefont {Esterlis}, \citenamefont {Volkov},
  \citenamefont {Scuri}, \citenamefont {Zhou}, \citenamefont {Brutschea},
  \citenamefont {Taniguchi}, \citenamefont {Watanabe}, \citenamefont {Yang},
  \citenamefont {Morales}, \citenamefont {Zhang}, \citenamefont {Millis},
  \citenamefont {Lukin}, \citenamefont {Kim}, \citenamefont {Demler},\ and\
  \citenamefont {Park}}]{sung2023}%
  \BibitemOpen
  \bibfield  {author} {\bibinfo {author} {\bibfnamefont {J.}~\bibnamefont
  {Sung}}, \bibinfo {author} {\bibfnamefont {J.}~\bibnamefont {Wang}}, \bibinfo
  {author} {\bibfnamefont {I.}~\bibnamefont {Esterlis}}, \bibinfo {author}
  {\bibfnamefont {P.~A.}\ \bibnamefont {Volkov}}, \bibinfo {author}
  {\bibfnamefont {G.}~\bibnamefont {Scuri}}, \bibinfo {author} {\bibfnamefont
  {Y.}~\bibnamefont {Zhou}}, \bibinfo {author} {\bibfnamefont {E.}~\bibnamefont
  {Brutschea}}, \bibinfo {author} {\bibfnamefont {T.}~\bibnamefont
  {Taniguchi}}, \bibinfo {author} {\bibfnamefont {K.}~\bibnamefont {Watanabe}},
  \bibinfo {author} {\bibfnamefont {Y.}~\bibnamefont {Yang}}, \bibinfo {author}
  {\bibfnamefont {M.~A.}\ \bibnamefont {Morales}}, \bibinfo {author}
  {\bibfnamefont {S.}~\bibnamefont {Zhang}}, \bibinfo {author} {\bibfnamefont
  {A.~J.}\ \bibnamefont {Millis}}, \bibinfo {author} {\bibfnamefont {M.~D.}\
  \bibnamefont {Lukin}}, \bibinfo {author} {\bibfnamefont {P.}~\bibnamefont
  {Kim}}, \bibinfo {author} {\bibfnamefont {E.}~\bibnamefont {Demler}},\ and\
  \bibinfo {author} {\bibfnamefont {H.}~\bibnamefont {Park}},\ }\href
  {https://arxiv.org/abs/2311.18069} {\bibinfo {title} {Observation of an
  electronic microemulsion phase emerging from a quantum crystal-to-liquid
  transition}} (\bibinfo {year} {2023}),\ \Eprint
  {https://arxiv.org/abs/2311.18069} {arXiv:2311.18069 [cond-mat.str-el]}
  \BibitemShut {NoStop}%
\bibitem [{\citenamefont {Kim}\ \emph {et~al.}(2022)\citenamefont {Kim},
  \citenamefont {Murthy}, \citenamefont {Pandey},\ and\ \citenamefont
  {Kivelson}}]{kim2022}%
  \BibitemOpen
  \bibfield  {author} {\bibinfo {author} {\bibfnamefont {K.-S.}\ \bibnamefont
  {Kim}}, \bibinfo {author} {\bibfnamefont {C.}~\bibnamefont {Murthy}},
  \bibinfo {author} {\bibfnamefont {A.}~\bibnamefont {Pandey}},\ and\ \bibinfo
  {author} {\bibfnamefont {S.~A.}\ \bibnamefont {Kivelson}},\ }\bibfield
  {title} {\bibinfo {title} {Interstitial-induced ferromagnetism in a
  two-dimensional wigner crystal},\ }\href
  {https://doi.org/10.1103/PhysRevLett.129.227202} {\bibfield  {journal}
  {\bibinfo  {journal} {Phys. Rev. Lett.}\ }\textbf {\bibinfo {volume} {129}},\
  \bibinfo {pages} {227202} (\bibinfo {year} {2022})}\BibitemShut {NoStop}%
\bibitem [{Note6()}]{Note6}%
  \BibitemOpen
  \bibinfo {note} {The importance of kinetic magnetism in triangular lattice
  Hubbard models has also been recently emphasized \cite
  {ciorciaro2023,*morera2024}.}\BibitemShut {Stop}%
\bibitem [{\citenamefont {{Castaing, B.}}(1980)}]{castaing1980}%
  \BibitemOpen
  \bibfield  {author} {\bibinfo {author} {\bibnamefont {{Castaing, B.}}},\
  }\bibfield  {title} {\bibinfo {title} {A model for exchange in liquid 3he},\
  }\href {https://doi.org/10.1051/jphyslet:019800041014033300} {\bibfield
  {journal} {\bibinfo  {journal} {J. Physique Lett.}\ }\textbf {\bibinfo
  {volume} {41}},\ \bibinfo {pages} {333} (\bibinfo {year} {1980})}\BibitemShut
  {NoStop}%
\bibitem [{\citenamefont {\ifmmode~\check{S}\else \v{S}\fi{}amaj}\ and\
  \citenamefont {Trizac}(2012)}]{samaj2012}%
  \BibitemOpen
  \bibfield  {author} {\bibinfo {author} {\bibfnamefont {L.}~\bibnamefont
  {\ifmmode~\check{S}\else \v{S}\fi{}amaj}}\ and\ \bibinfo {author}
  {\bibfnamefont {E.}~\bibnamefont {Trizac}},\ }\bibfield  {title} {\bibinfo
  {title} {Critical phenomena and phase sequence in a classical bilayer wigner
  crystal at zero temperature},\ }\href
  {https://doi.org/10.1103/PhysRevB.85.205131} {\bibfield  {journal} {\bibinfo
  {journal} {Phys. Rev. B}\ }\textbf {\bibinfo {volume} {85}},\ \bibinfo
  {pages} {205131} (\bibinfo {year} {2012})}\BibitemShut {NoStop}%
\bibitem [{\citenamefont {Xu}\ and\ \citenamefont {Balents}(2011)}]{Xu2011}%
  \BibitemOpen
  \bibfield  {author} {\bibinfo {author} {\bibfnamefont {C.}~\bibnamefont
  {Xu}}\ and\ \bibinfo {author} {\bibfnamefont {L.}~\bibnamefont {Balents}},\
  }\bibfield  {title} {\bibinfo {title} {Quantum phase transitions around the
  staggered valence-bond solid},\ }\href
  {https://doi.org/10.1103/PhysRevB.84.014402} {\bibfield  {journal} {\bibinfo
  {journal} {Phys. Rev. B}\ }\textbf {\bibinfo {volume} {84}},\ \bibinfo
  {pages} {014402} (\bibinfo {year} {2011})}\BibitemShut {NoStop}%
\bibitem [{\citenamefont {Sen}\ and\ \citenamefont {Sandvik}(2010)}]{Sen2010}%
  \BibitemOpen
  \bibfield  {author} {\bibinfo {author} {\bibfnamefont {A.}~\bibnamefont
  {Sen}}\ and\ \bibinfo {author} {\bibfnamefont {A.~W.}\ \bibnamefont
  {Sandvik}},\ }\bibfield  {title} {\bibinfo {title} {Example of a first-order
  n\'eel to valence-bond-solid transition in two dimensions},\ }\href
  {https://doi.org/10.1103/PhysRevB.82.174428} {\bibfield  {journal} {\bibinfo
  {journal} {Phys. Rev. B}\ }\textbf {\bibinfo {volume} {82}},\ \bibinfo
  {pages} {174428} (\bibinfo {year} {2010})}\BibitemShut {NoStop}%
\bibitem [{\citenamefont {Ciorciaro}\ \emph {et~al.}(2023)\citenamefont
  {Ciorciaro}, \citenamefont {Smole{\'n}ski}, \citenamefont {Morera},
  \citenamefont {Kiper}, \citenamefont {Hiestand}, \citenamefont {Kroner},
  \citenamefont {Zhang}, \citenamefont {Watanabe}, \citenamefont {Taniguchi},
  \citenamefont {Demler} \emph {et~al.}}]{ciorciaro2023}%
  \BibitemOpen
  \bibfield  {author} {\bibinfo {author} {\bibfnamefont {L.}~\bibnamefont
  {Ciorciaro}}, \bibinfo {author} {\bibfnamefont {T.}~\bibnamefont
  {Smole{\'n}ski}}, \bibinfo {author} {\bibfnamefont {I.}~\bibnamefont
  {Morera}}, \bibinfo {author} {\bibfnamefont {N.}~\bibnamefont {Kiper}},
  \bibinfo {author} {\bibfnamefont {S.}~\bibnamefont {Hiestand}}, \bibinfo
  {author} {\bibfnamefont {M.}~\bibnamefont {Kroner}}, \bibinfo {author}
  {\bibfnamefont {Y.}~\bibnamefont {Zhang}}, \bibinfo {author} {\bibfnamefont
  {K.}~\bibnamefont {Watanabe}}, \bibinfo {author} {\bibfnamefont
  {T.}~\bibnamefont {Taniguchi}}, \bibinfo {author} {\bibfnamefont
  {E.}~\bibnamefont {Demler}}, \emph {et~al.},\ }\bibfield  {title} {\bibinfo
  {title} {Kinetic magnetism in triangular moir{\'e} materials},\ }\href@noop
  {} {\bibfield  {journal} {\bibinfo  {journal} {Nature}\ }\textbf {\bibinfo
  {volume} {623}},\ \bibinfo {pages} {509} (\bibinfo {year}
  {2023})}\BibitemShut {NoStop}%
\bibitem [{\citenamefont {Morera}\ and\ \citenamefont
  {Demler}(2024)}]{morera2024}%
  \BibitemOpen
  \bibfield  {author} {\bibinfo {author} {\bibfnamefont {I.}~\bibnamefont
  {Morera}}\ and\ \bibinfo {author} {\bibfnamefont {E.}~\bibnamefont
  {Demler}},\ }\bibfield  {title} {\bibinfo {title} {{Itinerant magnetism and
  magnetic polarons in the triangular lattice Hubbard model}},\ }\href@noop {}
  {\bibfield  {journal} {\bibinfo  {journal} {arXiv preprint arXiv:2402.14074}\
  } (\bibinfo {year} {2024})}\BibitemShut {NoStop}%
\end{thebibliography}%

\end{document}